\documentclass[twocolumn]{aastex63}
\usepackage{bm}
\usepackage{threeparttable}
\usepackage{morefloats}
\usepackage{CJK}
\usepackage{amsmath}
\usepackage{enumerate}
\usepackage[shortlabels]{enumitem}

\def\bfnabla{{\mbox{\boldmath $\nabla$}}}

\renewcommand\bv{{\mbox{\boldmath $v$}}}
\newcommand\bb{{\mbox{\boldmath $B$}}}
\newcommand\bP{{\mbox{\boldmath $P$}}}
\newcommand\bn{{\mbox{\boldmath $n$}}}
\newcommand\bH{{\mbox{\boldmath $H$}}}

\newcommand\bF{{\mbox{\boldmath $F$}}}

\newcommand\bI{{{\sf\boldmath I}}}

\newcommand\Crat{{\mathbb{C}}}
\newcommand\Prat{{\mathbb{P}}}
\def\<{\,\langle\langle}
\def\>{\,\rangle\rangle}

\shorttitle{Implicit RT Algorithm}
\shortauthors{Jiang}

\begin{document}
\begin{CJK*}{UTF8}{gbsn}

\title{An Implicit Finite Volume Scheme to Solve the Time Dependent Radiation Transport Equation Based on Discrete Ordinates}

\correspondingauthor{Yan-Fei Jiang}
\email{yjiang@flatironinstitute.org}

\author[0000-0002-2624-3399]{Yan-Fei Jiang(姜燕飞)}
\affiliation{Center for Computational Astrophysics, \\
Flatiron Institute, \\
New York, NY 10010, USA}

\begin{abstract}
We describe a new algorithm to solve the time dependent, frequency integrated radiation transport (RT) equation implicitly, which is coupled to 
an explicit solver for equations of magnetohydrodynamics (MHD) using {\sf Athena++}. 
The radiation filed is represented by specific intensities along discrete rays, which 
are evolved using a conservative finite volume approach for both cartesian and curvilinear coordinate systems. 
All the terms for spatial transport of photons and interactions between 
gas and radiation are calculated implicitly together. An efficient Jacobi-like iteration scheme is used to solve the implicit equations. 
This removes any time step constrain due to the speed of light in RT. 
We evolve the specific intensities in the lab frame to simplify the transport step. The lab-frame specific intensities 
are transformed to the co-moving frame via Lorentz transformation when the source term is calculated. Therefore, the scheme does not 
need any expansion in terms of $v/c$. The radiation energy and momentum source terms for the gas are calculated via direct quadrature 
in the angular space. The time step for the whole scheme is determined by the normal Courant---Friedrichs---Lewy condition in the MHD module. 
We provide a variety of test problems for this algorithm including both optically thick and thin regimes, and for both gas 
and radiation pressure dominated flows to demonstrate its accuracy and efficiency. 

\end{abstract}


\section{Introduction}
\label{sec:introduction}
Thermal properties of the gas are determined by emission, absorption and scattering of photons as well as 
transport of the radiation field through the gas. In certain conditions, for example the flow around compact objects and in massive stars, the intense radiation field can be the dominant source of pressure force via the momentum exchange between photons and gas, which controls the dynamics of the system. Solving the radiation transport (RT) equation together with magneto-hydrodynamic (MHD) simulations becomes a very important task in numerical simulations of various astrophysical systems \citep[and references therein]{Teyssier2015}. 

RT equation is typically much more expensive to solve compared with the cost of  MHD simulations. The specific intensity, which is the fundamental quantity to describe the radiation field, is a function of time, spatial locations, propagating directions and frequency. Even adopting the gray approximation (frequency independent), resolving the angular distribution of photons is still harder to do compared with  MHD. In addition, photons travel with the speed of light in the optically thin regime, which is much faster than the flow speed of many non-relativistic systems. 
While in the very optically thick limit, photon mean free path can be much smaller than any other length scale of interest. Capturing different limits of the RT equation self-consistently while still evolving the MHD equations efficiently can be challenging. For these reasons, different assumptions have been made to simplify the RT equation. One type of the simplified approach is to take the moments of specific intensity and make certain assumptions to close the moment equations. Flux-limited diffusion (FLD) is a widely used method \citep[e.g.,][]{LevermorePomraning1981,TurnerStone2001,Krumholzetal2007,Holstetal2011}, which assumes that the mean radiation energy density follows the diffusion equation with a modified diffusion coefficient to limit the effective photon speed to be smaller than the speed of light. This assumption is well known to have issues in the optically thin regime. But more important, in the regime with intermediate optical depth when it is unclear whether diffusion approximation can be made or not, simulations adopting the FLD approximation can result in misleading conclusions \citep{KrumholzThompson2013}, which will not be known until the results are compared with simulations with more accurate methods \citep{Davisetal2014,TsangMilosavljevic2018,Smithetal2020}. Another increasingly popular alternative is the M1 method \citep{DubrocaFeugeas1999,Ponsetal2000,Gonzalezetal2007,SkinnerOstriker2013,Sadowskietal2013,McKinneyetal2013,Skinneretal2019}, which evolves the zeroth and first moments of specific intensity with an assumed closure relation based on local radiation energy density and flux. It is often claimed that the M1 approach is accurate in both optically thin and optically thick regimes because it can handle the shadow cast by one beam of radiation field. This is definitely incorrect as the assumed closure relation in M1 can easily fail in other cases. For example, it will merge multiple beams of photons into one since radiation field is treated as another fluid in this method. 

Other methods have been developed to solve the equation for specific intensities directly without assuming any closure relation. Monte Carlo based RT approach has been explored widely \citep[and references therein]{Whitney2011}, which is able to handle various radiative process accurately. Some novel schemes have also been developed to reduce the noise and computational cost of this type of method in different regimes \citep{Densmoreetal2007,Steinackeretal2013,RothKasen2015,TsangMilosavljevic2018,Foucart2018,Smithetal2020,RyanJoshua2020}. Another type of widely used method is based on discrete ordinates, or $S_n$, which solves the transport equation along discrete angles. Particularly, formal solution to the RT equation can be used as a closure to the radiation moment equation so that it can remove the limitations of FLD and M1. This variable Eddington tensor (VET) scheme has been used for a wide range of applications \citep{Stoneetal1992,HayesNorman2003,GehmeyrMihalas1994,Jiangetal2012,Davisetal2012,Asahinaetal2020}. The key of this approach is that the full time dependent evolution of radiation MHD is handled by the radiation moment equation, while the closure is given by the formal solution to the \emph{time independent} transport equation. Even though the RT equation is solved twice in different ways, it is often assumed that it will still be more efficient compared with solving the full time dependent RT equation directly. In this paper, we show that it is actually possible to solve the full time dependent RT equation in a finite volume approach efficiently with the computational cost comparable or even better than VET in some cases. The finite volume approach also makes it much easier to use in curvlinear coordinates, which can be complicated for the VET scheme.


The transport term in the RT equation is easier to solve in the laboratory frame 
while the interaction terms between radiation and gas are greatly simplified in 
the co-moving frame of the fluid. For non-relativistic systems where the flow velocity $v$ 
is much smaller than the speed of light $c$, RT equation is typically simplified by 
transforming the source terms from the co-moving frame to the laboratory frame only to certain 
orders of $v/c$ \citep{MihalasMihalas1984,MihalasKlein1982}. However, it has also been pointed out 
by \cite{Lowrieetal1999} that expansion only to the first order of $v/c$ is not enough. Some second order 
$v/c$ terms are needed to get the correct asymptotic limit in different regimes, particularly the dynamic diffusion 
regime \citep{Krumholzetal2007}. We want to point out that these expansions can be avoided completely if the 
specific intensities are solved directly. We can apply the Lorentz transformation to connect the radiation source terms 
between different frames. Even for non-relativistic systems, this approach is simple, accurate for all regimes 
and does not cause too much additional computational cost or complexity. 

When the typical flow speed in the system is close to speed of light, for example around compact objects, 
it is more efficient to solve the transport term of the RT equation explicitly by limiting the time step based on the speed of light 
\citep{Jiangetal2014,Sadowskietal2013,McKinneyetal2013,AnninosFragile2020,Asahinaetal2020}. However, for a wide range of 
applications with non-relativistic systems, it is necessary to solve the MHD equation explicitly while the RT equation is solved 
implicitly so that the time step is not limited by the speed of light. This is the approach adopted by most algorithms when the 
radiation moment equations are solved. When specific intensities are solved directly, the time dependent term is typically neglected. 
There have been several attempts to solve the full time dependent RT equations implicitly with mixed results \citep{StoneMihalas1992,SumiyoshiYamada2012}. 
When the time step is much larger than the light cross time per cell, the challenge is to keep a good balance between the transport term and the source term 
in all the asymptotic regimes. This is because the RT equation is a simple advection equation in the optically thin regime and it becomes a diffusion equation in the optically 
thick case. Advection will cause additional complexity. In this paper, we propose a fully implicit scheme that can get all the regimes correctly.

This paper is organized as follows.  In section \ref{sec:eq}, we describe the equations to solve, particularly on how the Lorentz transformation is used for the 
velocity dependent terms. The implicit discretization of the RT equation and detailed steps of the algorithm are described in section \ref{sec:num}. In section \ref{sec:test}, we show a set of test problems for our algorithm. Comparisons between our algorithm and some existing RT scheme as well as future development are discussed in section \ref{sec:discuss}.

\section{Equations}
\label{sec:eq}
The radiation MHD equations we solve are similar to what are used by \cite{Jiangetal2014}. However, 
we do not expand the source terms in orders of $v/c$. We will describe the RT equations and how they are 
coupled to the MHD equations in the following subsections. 

\subsection{ Equations for Radiation Transport}
We describe the radiation field using frequency integrated specific intensity $I$ in the lab frame, 
which is a function of time $t$, spatial location $(x,y,z)$ and angular direction as defined by the unit vector 
$\bn$. The transport 
term is greatly simplified in the lab frame while the source terms describing the interactions between radiation 
and gas are convenient in the co-moving frame of the fluid. Therefore, we also adopt this mixed frame approach 
\citep{Lowrieetal1999,Jiangetal2012,Jiangetal2014}. The RT equation we solve is (\citealt{MihalasKlein1982,MihalasMihalas1984})
\begin{eqnarray}
\frac{\partial I}{\partial t}+c\bn\cdot\bfnabla I=c\left(\eta-\chi I\right),
\end{eqnarray}
where $c$ is the speed of light. The emissivity $\eta$ and opacity $\chi$ are formally defined in the lab frame in the above equation. 
But the whole source term is calculated in the co-moving frame via the approach below. 
Traditionally, in order to get the expression for the source term, expansion to the first order 
of $\mathcal{O}\left(v/c\right)$ is usually used \citep{MihalasKlein1982}, with some additional 
second order terms \citep{Lowrieetal1999,Krumholzetal2007,Jiangetal2014}. Here we apply Lorentz transformation 
to the specific intensities to avoid the need of expansion in terms of $v/c$ completely. Although this may not 
sound necessary for non-relativistic systems, this approach does not cause additional computational cost and it results in 
a much cleaner calculation of the source terms. 

The frequency integrated specific intensities in the lab frame $I(\bn)$ is related to the co-moving frame values $I_0(\bn^{\prime})$ via 
the Lorentz transformation as \citep{MihalasMihalas1984}
\begin{eqnarray}
I_0(\bn^{\prime})=\gamma^4\left(1-\bn\cdot\bv/c\right)^4 I(\bn)\equiv \Gamma^4(\bn,\bv)I(\bn),
\label{eq:lorentz_transform}
\end{eqnarray} 
where $\bv$ is the flow velocity, $\gamma\equiv 1/\sqrt{1-v^2/c^2}$ is the corresponding Lorentz factor and $\bn^{\prime}$ 
is the angle in the co-moving frame given by
\begin{eqnarray}
\bn^{\prime}=\frac{1}{\gamma\left(1-\bn\cdot\bv/c\right)}\left[\bn-\gamma\frac{\bv}{c}\left(1-\frac{\gamma}{\gamma+1}\frac{\bn\cdot\bv}{c}\right)\right].
\label{eqn:cm_ang}
\end{eqnarray}
We have defined $\Gamma(\bn,\bv)\equiv \gamma\left(1-\bn\cdot\bv/c\right)$ for convenience. We multiply the left hand sides of the 
source term with $\Gamma^4$ to get
\begin{eqnarray}
\frac{\partial I_0}{\partial t}=c\Gamma \left[\eta_0 - \chi_0 I_0\right].
\end{eqnarray}
Here the frequency integrated emissivity $(\eta_0)$ and opacity $(\chi_0)$ in the co-moving frame 
are related to $\eta,\chi$ as
\begin{eqnarray}
\eta_0=\Gamma^3\eta,~~ \chi_0=\Gamma^{-1}\chi.
\end{eqnarray}
Under the assumption of local thermal equilibrium, we can write the frequency integrated source term in the co-moving frame as
\begin{eqnarray}
\frac{\partial I_0}{\partial t}&=&c\Gamma\left[\rho\kappa_s \left(J_0 - I_0\right) \right. \nonumber\\
&+& \left.  \rho\kappa_a\left(\frac{a_rT^4}{4\pi}-I_0\right) 
 + \rho\kappa_{\delta P}\left(\frac{a_rT^4}{4\pi}-J_0\right)\right]\nonumber\\
 &=&c\Gamma\left[\rho(\kappa_s+\kappa_a)\left(J_0-I_0\right)  \right. \nonumber\\
&+& \left. \rho\left(\kappa_a+\kappa_{\delta P}\right)\left(\frac{a_rT^4}{4\pi} - J_0\right)
 \right],
 \label{eq:cm_source}
\end{eqnarray}
and the corresponding equation for gas internal energy $E_g$:
\begin{eqnarray}
\frac{\partial E_g}{\partial t}=-c\rho\left(\kappa_a+\kappa_{\delta P}\right)\left(a_rT^4-4\pi J_0\right).
\end{eqnarray}
Here $a_r$ is the radiation constant and $\rho,~ T$ are gas density and temperature, which should be co-moving frame quantities in principle. When the RT algorithm is coupled to a Newtonian MHD solver,
we do not distinguish gas quantities as well as time in the lab and co-moving frames. The angular averaged mean intensity in the co-moving frame $J_0$ is defined as
\begin{eqnarray}
J_0\equiv \frac{1}{4\pi}\int I_0d \Omega_0,
\end{eqnarray}
where $\Omega_0$ is the angular element in the co-moving frame. The specific scattering opacity $\kappa_s$ (with unit of $cm^2/g$) is assumed to be isotropic and does not change 
the mean photon energy such as electron scattering. The Rosseland mean absorption opacity is $\kappa_a$ and we take $\kappa_{\delta P}$ to be the difference between the Planck mean and Rosseland mean opacity. This is designed to ensure that when we integrate the zeroth and first moment of the co-moving frame specific intensity over the angular space, we get the energy and momentum source terms $\rho\left(\kappa_a+\kappa_{\delta P}\right)\left(a_rT^4-4\pi J_0\right)$ and $-\rho\left(\kappa_s+\kappa_a\right)\left(4\pi \bH_0\right)$ 
with $\bH_0\equiv \int I_0\bn^{\prime} d\Omega_0/\left(4\pi\right)$. This is consistent with the Rosseland and Planck mean opacities normally used for radiation moment equations. 
We multiply both sides of equation \ref{eq:cm_source} with  $\Gamma^{-4}$ to get the source term for lab-frame specific intensities. 

\subsection{Equations of Radiation MHD}
Photons are coupled to the gas via momentum and energy exchange. Since the total energy and momentum are 
conserved in the lab frame, we integrate the source terms for lab frame specific intensities over the angles to get the full 
radiation MHD equations as
\begin{align}
\frac{\partial\rho}{\partial t}+\bfnabla\cdot(\rho \bv)&=0, \nonumber \\
\frac{\partial( \rho\bv)}{\partial t}+\bfnabla\cdot({\rho \bv\bv-\bb\bb+{{\sf P}^{\ast}}}) &=-\bm{ S_r}(\bP),\  \nonumber \\
\frac{\partial{E}}{\partial t}+\bfnabla\cdot\left[(E+P^{\ast})\bv-\bb(\bb\cdot\bv)\right]&=-S_r(E),  \nonumber \\
\frac{\partial\bb}{\partial t}-\bfnabla\times(\bv\times\bb)&=0, \nonumber\\
\frac{\partial I}{\partial t}+c\bn\cdot\bfnabla I&=cS_I,\nonumber\\
S_I\equiv \Gamma^{-3}\left[\right. \rho(\kappa_s + \kappa_a)\left(J_0-I_0\right)\nonumber\\
+\rho(\kappa_a+\kappa_{\delta P})\left(\frac{a_rT^4}{4\pi}-J_0\right)\left.\right],\nonumber\\
S_r(E)\equiv 4\pi c\int S_I d\Omega,\nonumber\\
\bm{S_r}(\bP)\equiv 4\pi \int \bn S_I d\Omega.
\label{eq:mhd}
\end{align}
Here $\bb$ is the magnetic field, ${\sf P}^{\ast}\equiv(P+B^2/2){\sf I}$ (with ${\sf I}$
the unit tensor), and the magnetic permeability $\mu$ is taken to be $1$ in this unit system. 
The total gas energy density is
\begin{eqnarray}
E=E_g+\frac{1}{2}\rho v^2+\frac{B^2}{2}.
\end{eqnarray}
We assume ideal gas so that the gas internal 
energy is related to gas pressure via the adiabatic index $\gamma_g$ as 
$E_g=P/(\gamma_g-1)$ for $\gamma_g \neq 1$. The gas temperature is calculated with 
$T=P/\left(R_{\text{ideal}}\rho\right)$, where $R_{\text{ideal}}$ is the ideal gas constant.

\section{The Implicit Solver for the RT Equation}
\label{sec:num}
We will use the normal explicit Godunov scheme with constrained transport in {\sf Athena++} 
to evolve the MHD part of equations \ref{eq:mhd}. The time step is limited by the CFL condition 
based on the maximum flow velocity, sound speed and Alfv\'en speed. For the RT equation, 
we will solve all the terms implicitly to remove the time step constrain due to speed of light. 
We will describe how we discretize and solve the RT equation in this section. 

\subsection{Spatial and Angular Discretization}
\label{sec:spatial_angles}
The radiation field shares the same spatial grid as used for the gas quantities in {\sf Athena++}. 
We label the three coordinates as $x,y,z$, which can be any orthogonal curvilinear systems such as cartesian, 
cylindrical or spherical polar coordinates. The total number of grid points along the three directions are $N_x, N_y, N_z$ 
respectively. Notice that our finite volume scheme does not need the grid spacing to be uniform. What we need is 
the volume of each cell $V_{i,j,k}$ for $i\in [0,N_x-1], j\in [0,N_y-1], k\in [0,N_z-1]$ as well as surface areas of each cell 
facing the three directions $A_x, A_y, A_z$. All the specific intensities are defined at the volume center of each cell. 

We adopt the same angular discretization scheme as used by \cite{Davisetal2012} and \cite{Jiangetal2014} to be our 
default choice\footnote{Other angular discretization approaches can also be used for specific applications as shown in 
Section \ref{sec:sphere_angles}}. The angular unit vectors $\bn$ are defined with respect to a global cartesian coordinate and they do 
not change with spatial locations.  The three components of each angle $\bn=\left(\mu_x,\mu_y,\mu_z\right)$, as well 
as the quadrature weight $w_n$ are determined based on the same algorithm developed by \cite{Brulsetal1999} 
following the original quadrature principle described in \cite{Carlson1963}. This set of angles ensures that the following 
relations are satisfied numerically
\begin{align}
\sum_{n=0}^{N-1} w_n &= 1,\nonumber\\
\sum_{n=0}^{N-1} \mu_x w_n &= \sum_{n=0}^{N-1} \mu_y w_n = \sum_{n=0}^{N-1} \mu_z w_n = 0, \nonumber\\
\sum_{n=0}^{N-1} \mu^2_x w_n &= \sum_{n=0}^{N-1} \mu_y^2 w_n = \sum_{n=0}^{N-1} \mu^2_z w_n = \frac{1}{3}, 
\end{align}
where $N$ is the total number of angles. 

Notice that these angles are always defined in the lab frame. The angles in 
the co-moving frame are determined based on equation \ref{eqn:cm_ang} while the quadrature weight in the co-moving frame is 
$w_n^{\prime}=\Gamma^{-2}w_n$.
For infinite number of angles, it can be shown that $\int \Gamma^{-2} d\Omega_0=1$ for any flow velocity. However, this relation 
normally is not satisfied numerically to machine precision due to truncation errors with a finite number of angles. We correct this error 
by normalizing $w_n^{\prime}$ as
\begin{eqnarray}
w^{\prime}_n\equiv \frac{\Gamma^{-2}w_n}{\sum_{n=0}^{N-1}\Gamma^{-2}w_n}.
\end{eqnarray}
Therefore $\sum_0^{N-1}w_n^{\prime}=1$ is always satisfied. We find it is necessary to correct this angular truncation error 
to avoid accumulative energy error when flow velocity is not zero.  The commonly used radiation energy density, radiation flux and 
radiation pressure in the lab and co-moving frames are just derived quantities from specific intensities as 
\begin{align}
E_r=4\pi \sum_{n=0}^{N-1} I_n w_n&,E_{r,0}=4\pi \sum_{n=0}^{N-1} I_{0,n} w_n^{\prime},\nonumber\\
\bF_r=4\pi c\sum_{n=0}^{N-1} I_n \bn w_n&,\bF_{r,0}=4\pi c\sum_{n=0}^{N-1} I_{0,n} \bn^{\prime} w_n^{\prime},\nonumber\\
{\sf P}_r=4\pi \sum_{n=0}^{N-1} I_n \bn\bn w_n&,{\sf P}_{r,0}=4\pi \sum_{n=0}^{N-1} I_{0,n} \bn^{\prime}\bn^{\prime} w_n^{\prime}.
\end{align}

\subsection{Discretized Equation for RT}
The radiation energy source term can be very stiff in the optically thick regions while the radiation momentum source term is typically not \citep{SekoraStone2010,Jiangetal2012}. Therefore, when we solve the RT equation, we will assume the flow velocity and gas density do not change. But we evolve the gas temperature with the radiation variables together to ensure that correct thermal equilibrium state is achieved. As higher order spatial reconstruction without flux limiter cannot guarantee total variation diminishing, which is necessary to avoid oscillatory behavior around local maximums or minimums, we will only use first order spatial reconstruction for our algorithm. The commonly used flux limiters will make the transport term non-linear, which is very hard to solve implicitly. The algorithm will also be first order accurate in time for RT to avoid oscillatory behavior during the temporal evolution of gas temperature and radiation energy density when the source term is very stiff. 

For given initial conditions $I_n^m, \rho^m, T^m, v^m$ at time step $m$, the specific intensities $I_n^{m+1}$ with $n\in[0,N-1]$ at time step $m+1$ are calculated based on the following 
$N+1$ equations
\begin{align}
I^{m+1}_n-I_n^m+\Delta t c\bn\cdot\bfnabla I_n^{m+1}=\nonumber\\
\Delta t c\Gamma_n^{-3}\left[\right.\rho^m\kappa_s\left(J_0^{m+1}-I_{0,n}^{m+1}\right)\nonumber\\
+\rho^m\kappa_a\left(\frac{a_r\left(T^{m+1}\right)^4}{4\pi}-I_{0,n}^{m+1}\right)\nonumber\\
+\rho^m\kappa_{\delta P}\left(\frac{a_r(T^{m+1})^4}{4\pi}-J_0^{m+1}\right)&\left.\right],\nonumber\\
\frac{\rho^m R_{\text{ideal}}}{\gamma_g-1}\left(T^{m+1}-T^m\right)=\nonumber\\
-\Delta t c\rho^{m}\left(\kappa_a+\kappa_{\delta P}\right)\left[a_r(T^{m+1})^4-4\pi J_0^{m+1}\right] ,
\label{eqn:In}
\end{align}
where 
\begin{eqnarray}
I_{0,n}^{m+1}=\Gamma^4_n I_n^{m+1},J_0^{m+1}=\sum_{n=0}^{N-1}w_n^{\prime} I_{0,n}^{m+1},
\end{eqnarray}
and $\Gamma_n$ is calculated using $v^m$ for each angle. The opacities ($\kappa_s, \kappa_a$ and $\kappa_{\delta P}$) can be 
functions of local gas properties in principle. But we will assume they keep the values at time step $m$ in the above equations. 
This is a set of fully coupled equations for gas temperature and specific intensities at all angles (via the terms with $J_0^{m+1}$ 
and spatial locations (via the transport term $\bn\cdot\bfnabla I_n^{m+1}$). These equations are only linear with respect to $I_n^{m+1}$ 
but not for $T^{m+1}$ due to the $(T^{m+1})^4$ term. All the transport and source terms are calculated using 
the variables at time step $m+1$ to make the scheme unconditional stable for any time step $\Delta t$. 
 We will use an iterative approach to solve these equations as described below. 

\subsubsection{The Transport Term}
\label{sec:transport}
Following \cite{Jiangetal2014}, we can rewrite the transport term in the above equations as 
$c\bfnabla\cdot\left(\bn I_n^{m+1}\right)$ with our default angular discretization 
because $\bn$ does not change with spatial locations. Then this term becomes the conservative format 
and can be calculated with the standard finite volume approach. 
For interface at $i-1/2, j, k$, we simply take the left and right states for $I_n^{m+1}$ to be $I_n^{m+1}(i-1,j,k)$ 
and $I_n^{m+1}(i,j,k)$ respectively. The same approach is used for other directions.

Special care is needed to determine the surface fluxes for each specific intensity. If we simply take the upwind fluxes as in \cite{StoneMihalas1992}, although total energy 
is conserved, numerical diffusion can overcome the physical diffusion in the very optically thick regime. This can be easily demonstrated with 
only two angles in one dimension. If we assume the two angles have $\mu_{x,1}> 0$ and $\mu_{x,2} < 0$, 
the upwind fluxes at $i-1/2$ will be $c\mu_{x,1}I_1(i-1)$ and $c\mu_{x,2}I_2(i)$. 
Similarly, the fluxes at $i+1/2$ for the two angles are $c\mu_{x,1}I_1(i)$ and $c\mu_{x,2}I_2(i+1)$. The change of the mean radiation energy density $w_1I_1+w_2I_2$ 
at cell $i$ is then determined by $cw_2\mu_{x,2}I_2(i+1)+cw_1\mu_{x,1}I_1(i)-cw_2\mu_{x,2}I_2(i)-cw_1\mu_{x,1}I_1(i-1)$. 
However, radiation moment equation requires that it should be proportional to $F_r(i+1)-F_r(i-1)=cw_2\mu_{x,2}I_2(i+1)+cw_1\mu_{x,1}I_1(i+1)-cw_2\mu_{x,2}I_2(i-1)-cw_1\mu_{x,1}I_1(i-1)$. The difference is essentially the truncation error, which is negligible in the optically thin regime. However, in the optically thick regime when 
the radiation field is very close to be isotropic, $|I_2(i+1)-I_1(i+1)|$ can be much smaller than $|I_1(i+1)-I_1(i)|$. This will cause the numerical 
flux due to the truncation errors to be much larger than the physical flux due to photon diffusion, which results in completely wrong numerical solutions. We have designed a modified HLLE flux to overcome the issue. 

The dynamical diffusion regime also needs to be treated carefully as discussed in \cite{Jiangetal2014}. The numerical 
truncation error due to finite flow velocity can also overcome the physical diffusion when the optical depth per cell is very large particularly with first order spatial reconstruction. 
In principle, these numerically errors can be minimized with second or third order spatial reconstructions. However, it is not an option for our implicit scheme. 
Instead, we find a specially designed HLLE type flux can work very well in all the regimes. 

We rewrite the transport term as 
\begin{eqnarray}
c\bfnabla\cdot\left(\bn I_n^{m+1}\right)=\bfnabla\cdot\left[\left(c\bn-f \bv^m\right)I_n^{m+1}\right]\nonumber\\
+\bfnabla\cdot\left(f\bv^m I_n^{m+1}\right).
\end{eqnarray}
Here $f$ is a function of optical depth per cell $\tau_c\equiv \rho^m\left(\kappa_a+\kappa_s\right)\Delta L$ with $\Delta L$ to be a constant factor 
times the cell size. The function is chosen as 
\begin{eqnarray}
f=1-\exp(-\tau_c^2).
\label{eq:taucell}
\end{eqnarray}
This is designed to separate the advection term when the photon mean free path is not resolved ($f\approx 1$). This is not necessary when 
the optical depth per cell is much smaller than 1 ($f\approx 0$). Since our time step will satisfy the CFL condition for flow velocity, we can calculate 
the advective term $\bfnabla\cdot\left(f\bv^{m} I^{m+1}\right)$ explicitly, which means 
\begin{eqnarray}
c\bfnabla\cdot\left(\bn I_n^{m+1}\right)\approx\bfnabla\cdot\left[\left(c\bn-f \bv^m\right)I_n^{m+1}\right]\nonumber\\
+\bfnabla\cdot\left(f\bv^m I_n^{m}\right).
\end{eqnarray}
We use second order spatial reconstruction for $I_n^{m}$ to get the values at the cell faces and then upwind flux based on $f\bv^m$ to calculate the 
term $\bfnabla\cdot\left(f\bv^m I_n^{m}\right)$. 

For the interface at $(i-1/2,j,k)$, we take the left and right states for each specific intensity to be $I_{n,L}=I^{m+1}_n(i-1,j,k)$ and $I_{n,R}=I^{m+1}_n(i,j,k)$. 
Then we calculate the flux $F_n(i-1/2)$ at the interface for each angle based on a modified HLLE formula (Appendix \ref{appendix:coef}). 
The key modification is to make sure the flux is the upwind value in the optically thin case and it gets close 
to the centered value in the optically thick case based on the parameter $\tau_c$. Similarly, we calculate the flux $F_n(i+1/2)$
for the interface $(i+1/2,j,k)$ as well as fluxes along other directions $F_n(j-1/2),F_n(j+1/2),F_n(k-1/2),F_n(k+1/2)$. 
Detailed expressions for the fluxes are  given in the Apendix \ref{appendix:coef}. Then the flux divergence term is calculated as
\begin{align}
&\bfnabla\cdot\left[\left(c\bn - f\bv^m\right) I_n^{m+1}\right]=\frac{1}{V_{i,j,k}}\nonumber\\
&\left[\right. A_x(i+1/2)F_n(i+1/2)-A_x(i-1/2)F_n(i-1/2)\nonumber\\
&+A_y(j+1/2)F_n(j+1/2)-A_y(j-1/2)F_n(j-1/2)\nonumber\\
&+A_z(k+1/2)F_n(k+1/2)-A_z(k-1/2)F_n(k-1/2)\left. \right].
\end{align} 
Reorganizing all the terms results in the following equation that we will apply our iterative solver
\begin{align}
&g_1 I^{m+1}_n+g_2I_n^{m+1}(i-1)+g_3I_n^{m+1}(i+1)\nonumber\\
+&g_4I_n^{m+1}(j-1)+g_5I_n^{m+1}(j+1)\nonumber\\
+&g_6I_n^{m+1}(k-1)+g_7I_n^{m+1}(k+1)\nonumber\\
=&I_n^m-\Delta t\bfnabla\cdot\left(f\bv^mI_n^m\right)\nonumber\\
+&\Delta t c\Gamma_n^{-3}\left[\right.\rho^m\kappa_s\left(J_0^{m+1}-I_{0,n}^{m+1}\right)\nonumber\\
+&\rho^m\kappa_a\left(\frac{a_r\left(T^{m+1}\right)^4}{4\pi}-I_{0,n}^{m+1}\right)\nonumber\\
+&\rho^m\kappa_{\delta P}\left(\frac{a_r(T^{m+1})^4}{4\pi}-J_0^{m+1}\right)\left.\right],\nonumber\\
&\frac{\rho^m R_{\text{ideal}}}{\gamma_g-1}\left(T^{m+1}-T^m\right)=\nonumber\\
-&\Delta t c\rho^{m}\left(\kappa_a+\kappa_{\delta P}\right)\left[a_r(T^{m+1})^4-4\pi J_0^{m+1}\right].
\label{eqn:In_final}
\end{align}
Expressions for the coefficients $g_1$ to $g_7$ are given in the Appendix \ref{appendix:coef}. 
Notice that their values depend on the local optical depth $\tau_c$ and they only need to be calculated once for each time step.

\subsubsection{The Iterative Solver}
We need to solve the $N+1$ equations (\ref{eqn:In_final}) simultaneously to get the solutions for 
$I_n^{m+1}$ ($n\in[0,N-1]$) and $T^{m+1}$. These equations are only linear for $I_n^{m+1}$. 
These terms can be rewritten as a sparse matrix with size of $N\times N_x\times N_y\times N_z$. 
If the specific intensities are ordered in a 1D array such that $I_n^{m+1}(i,j,k)$ is located at 
$kN_yN_xN+jN_xN+iN+n$, the matrix has the special structure that the diagonal direction are blocks with 
size $N\times N$, which couple all the angles at the same spatial locations. All the other elements are only 
non-zero at locations corresponding to $(n,i\pm 1,j,k)$, $(n,i,j\pm 1, k)$ and $(n,i,j,k\pm 1)$. The terms related 
to $T^{m+1}$ are non-linear but $T^{m+1}$ at different spatial locations are completely decoupled. 

This motivates us to solve these equations iteratively similar to the Jacobi approach used for linear systems.  
For each cell, we take all the specific intensities at different spatial locations to be the values from the last step. Then 
we go through each cell and solve the equations for all the specific intensities at once. We update the off-diagonal terms with the updated 
specific intensities and continue this process until changes of the specific intensities between neighboring steps are below a certain threshold. 

Specifically, we solve the following equations iteratively
\begin{align}
&g_1 I^{m+1}_{n,l}+g_2I_{n,l-1}^{m+1}(i-1)+g_3I_{n,l-1}^{m+1}(i+1)\nonumber\\
+&g_4I_{n,l-1}^{m+1}(j-1)+g_5I_{n,l-1}^{m+1}(j+1)\nonumber\\
+&g_6I_{n,l-1}^{m+1}(k-1)+g_7I_{n,l-1}^{m+1}(k+1)\nonumber\\
=&I_{n}^m-\Delta t\bfnabla\cdot\left(f\bv^mI_n^m\right)\nonumber\\
+&\Delta t c\Gamma_n^{-3}\left[\right.\rho^m\kappa_s\left(J_{0,l}^{m+1}-I_{0,n,l}^{m+1}\right)\nonumber\\
+&\rho^m\kappa_a\left(\frac{a_r\left(T^{m+1}_l\right)^4}{4\pi}-I_{0,n,l}^{m+1}\right)\nonumber\\
+&\rho^m\kappa_{\delta P}\left(\frac{a_r(T^{m+1}_l)^4}{4\pi}-J_{0,l}^{m+1}\right)\left.\right],\nonumber\\
&\frac{\rho^m R_{\text{ideal}}}{\gamma_g-1}\left(T^{m+1}_l-T^m\right)=\nonumber\\
-&\Delta t c\rho^{m}\left(\kappa_a+\kappa_{\delta P}\right)\left[a_r(T^{m+1}_l)^4-4\pi J_{0,l}^{m+1}\right],
\label{eqn:iteration}
\end{align}
where $l$ represents each step during the iterative process. At the beginning, we need an initial guess for specific intensities 
at the whole grid for all the angles. This is usually taken to be the solution from the last time step or the initial condition. Then we go through 
all the grid points to solve the above equation for all the angles at each cell together in the following way. 

For each step during the iteration, since all the variables $I_{n,l-1}^{m+1}(i-1),I_{n,l-1}^{m+1}(i+1),I_{n,l-1}^{m+1}(j-1),I_{n,l-1}^{m+1}(j+1)),
I_{n,l-1}^{m+1}(k-1),I_{n,l-1}^{m+1}(k+1)$ are already known, we can reorganize all the equations we need to solve at each cell as 
\begin{align}
g_1 I_{n,l}^{m+1}&=I_c+\Delta t c\Gamma_n^{-3}\left[\right.\rho^m\kappa_s\left(J_{0,l}^{m+1}-I_{0,n,l}^{m+1}\right)\nonumber\\
+&\rho^m\kappa_a\left(\frac{a_r\left(T^{m+1}_l\right)^4}{4\pi}-I_{0,n,l}^{m+1}\right)\nonumber\\
+&\rho^m\kappa_{\delta P}\left(\frac{a_r(T^{m+1}_l)^4}{4\pi}-J_{0,l}^{m+1}\right)\left.\right],
\label{eq:rad_source}
\end{align}
\begin{align}
&\frac{\rho^m R_{\text{ideal}}}{\gamma_g-1}\left(T^{m+1}_l-T^m\right)=\nonumber\\
-&\Delta t c\rho^{m}\left(\kappa_a+\kappa_{\delta P}\right)\left[a_r(T^{m+1}_l)^4-4\pi J_{0,l}^{m+1}\right],
\label{eq:gas_source}
\end{align}
where $I_c$ includes all the other terms that are already known before step $l$. We multiply the left and right sides of the equation for each $I_{n,l}^{m+1}$ with $\Gamma_n^4$ 
to get the equation for co-moving frame specific intensities as
\begin{align}
g_1 I_{0,n,l}^{m+1}&=\Gamma_n^4 I_c+\Delta t c\Gamma_n\left[\right.\rho^m\kappa_s\left(J_{0,l}^{m+1}-I_{0,n,l}^{m+1}\right)\nonumber\\
+&\rho^m\kappa_a\left(\frac{a_r\left(T^{m+1}_l\right)^4}{4\pi}-I_{0,n,l}^{m+1}\right)\nonumber\\
+&\rho^m\kappa_{\delta P}\left(\frac{a_r(T^{m+1}_l)^4}{4\pi}-J_{0,l}^{m+1}\right)\left.\right].
\label{eq:rad_source2}
\end{align}
We sum up  the equations with the appropriate weight $w_n^{\prime}$ for each angle and get
\begin{align}
g_1 J_{0,l}^{m+1}&=\sum_nw_n^{\prime} \Gamma_n^4 I_c+\Delta t c\Gamma_n\left[\right.\nonumber\\
+&\rho^m\kappa_a\left(\frac{a_r\left(T^{m+1}_l\right)^4}{4\pi}-J_{0,l}^{m+1}\right)\nonumber\\
+&\rho^m\kappa_{\delta P}\left(\frac{a_r(T^{m+1}_l)^4}{4\pi}-J_{0,l}^{m+1}\right)\left.\right].
\end{align}
The scattering term drops out because it does not change the mean photon energy. 
Together with equation \ref{eq:gas_source}, we can get the solution for $J_{0,l}^{m+1}$ and $T_l^{m+1}$ by performing a 
root finding for a fourth order polynomial. Then 
specific intensity for each angle $I_{0,n,l}^{m+1}$ can then be calculated from equation \ref{eq:rad_source2} independently, 
which can be converted to $I_{n,l}^{m+1}$ directly. In this way, the non-linearity of the source term, as well as  the coupling between 
different angles, all come down to a four order polynomial solver, which is trivial and independent of the number of angles. Therefore, 
the whole cost for each step during the iteration is linearly proportional to the number of angles. 
We continue the process for each grid and 
stop the iteration when 
\begin{equation}
\Delta I\equiv \sum_{n,i,j,k}|I^{m+1}_{n,l}-I^{m+1}_{n,l-1}|/\sum_{n,i,j,k}|I^{m+1}_{n,l}|
\label{eq:error}
\end{equation}
 is smaller than a threshold, typically $10^{-5}$.

\subsubsection{The Radiation Source Terms}
Since total energy and momentum are conserved in the lab frame, we use the conservation laws to determine the 
source terms we need to add to the gas. The radiation energy and momentum at the beginning of the time step is 
\begin{eqnarray}
E_r^m=4\pi \sum_{n=0}^{N-1}I^{m}_n w_n, 
\bF_r^{m}=4\pi c\sum_{n=0}^{N-1}\bn I^{m}_n w_n.
\end{eqnarray}
At the end of the time step, they can be calculated as 
\begin{align}
E_r^{m+1}&=4\pi \sum_{n=0}^{N-1}I^{m+1}_n w_n, \nonumber\\
\bF_r^{m+1}&=4\pi c\sum_{n=0}^{N-1}\bn I^{m+1}_n w_n.
\end{align}
Part of the change is due to the transport term, which can also be calculated as
\begin{align}
\Delta E_r&=4\pi \sum_{n=0}^{N-1} \left\{\right.\Delta t\bfnabla\cdot\left[\left(c\bn - f\bv^m\right) I_n^{m+1}\right] \nonumber \\
&+\Delta t\bfnabla\cdot\left(f\bv^mI_n^m\right) \left. \right\}w_n, \nonumber\\
\Delta \bF_r&=4\pi c\sum_{n=0}^{N-1} \left\{\right.\Delta t\bfnabla\cdot\left[\left(c\bn - f\bv^m\right) I_n^{m+1}\right] \nonumber \\
&+\Delta t\bfnabla\cdot\left(f\bv^mI_n^m\right) \left.\right\}\bn w_n.
\end{align}
The source terms are then given by
\begin{align}
S_r(E)&=E_r^{m+1} - E_r^m - \Delta E_r\nonumber\\
\bm{ S_r}(\bP)&=\frac{1}{c^2}\left(\bF_r^{m+1}-\bF_r^m-\Delta \bF_r\right).
\label{eqn:add_source}
\end{align}

\subsubsection{Additional Terms with Spherical Polar Angular System}
\label{sec:sphere_angles}
The choice of the angular system is not unique. If the angles vary with spatial locations, additional terms 
will show up when we convert $\bn\cdot\bfnabla I$ to $\bfnabla\cdot \left(\bn I\right)$. The RT equation 
with general angular systems is described in \cite{DavisGammie2020}. One special case that is useful for 
non-relativistic system is the angular system in spherical polar coordinate, which allows us to use this scheme in 1D or 2D spherical polar coordinates. 
Notice that our default angular system as described in Section \ref{sec:spatial_angles} can also be used in principle. The alternative angular 
system we describe here can be more convenient and efficient for certain applications. 

In spherical polar coordinate, we define a set of angles with respect to the local coordinate $(r,\theta,\phi)$ in each cell. We divide the longitudinal angles $\psi$ uniformly from 
$0$ to $2\pi$. For the polar angle $\zeta$, it is uniformly divided in $\cos\zeta$ for $\zeta\in[0,\pi]$. These angles point to different directions at different spatial locations, which introduce additional terms in the transport term as \citep{DavisGammie2020}
\begin{align}
\frac{\partial I}{\partial t} &+c\bn\cdot\bfnabla I=\frac{\partial I}{\partial t}+c\bfnabla\cdot\left(\bn I\right)\nonumber\\
&-\frac{1}{r\sin\zeta}\frac{\partial\left(\sin^2\zeta I\right)}{\partial \zeta}-\frac{\sin\zeta\cos\theta}{r\sin\theta}\frac{\partial\left(\sin\psi I\right)}{\partial \psi}.
\end{align} 
All the other terms are unchanged. The numerical scheme described in the previous sections remains the same except that we need to calculate the change of $I$ 
due to the new terms.

The new terms represent transport of specific intensities in the angular space, which we can calculate implicitly as
\begin{align}
&\frac{I^{m+1}_n-I^m_n}{\Delta t}\nonumber\\
-&\frac{1}{r}\frac{\sin^2\zeta_{n+1/2}I_{n+1/2}^{m+1} - \sin^2\zeta_{n-1/2}I_{n-1/2}^{m+1}}{\cos\zeta_{n-1/2}-\cos\zeta_{n+1/2}}\nonumber\\
-&\frac{\sin\zeta_n\cos\theta}{r\sin\theta}\frac{\sin\psi_{n+1/2}I_{n+1/2}^{m+1}-\sin\psi_{n-1/2}I_{n-1/2}^{m+1}}{\psi_{n+1/2}-\psi_{n-1/2}}=0.
\end{align}
Here the variables with subscripts $n-1/2$ and $n+1/2$ represent left and right hand sides in the angular space. We use first order upwind values for 
$I_{n-1/2}^{m+1}$ and $I_{n+1/2}^{m+1}$. During each iteration, we modify the coefficient $g_1$ in equation \ref{eqn:iteration} to include additional contributions  for $I_{n,l}^{m+1}$ from this term. For all the other specific intensities with different angles in the above equation, we use values from the last iteration, which are added to the right hand side of equation \ref{eqn:iteration}. We find it is necessary to include all the terms during the iterative process in order to maintain a balance between all the transport terms and source terms. 

\subsection{Algorithm for the Full Radiation MHD Equation}
The implicit solver is coupled to the two step van-Leer MHD solver in {\sf Athena++} \citep{Stoneetal2020}. 
We will not repeat  details of the Godunov scheme with constrained transport for MHD here. Instead, we just outline the steps of the full 
algorithm here. 
\begin{enumerate}[(a)]
\item Determine the time step $\Delta t$ based on the flow speed, sound speed, Alfv\'en speed and cell sizes with CFL number $0.4$.
\item Update the gas quantities and magnetic field for half time step $\Delta t/2$ with the MHD solver in {\sf Athena++}.
\item Evolve the radiation field for half time step $\Delta t/2$ as described in the previous section 
to determine the energy and momentum source terms 
$S_r(E), \bm{ S_r}(\bP)$ (equation \ref{eqn:add_source}). Add these terms to gas momentum and total energy. 
\item Apply the MHD solver for a full time step $\Delta t$ using the partially updated variables at $\Delta t/2$. 
\item Repeat the implicit solver of the radiation field for a full time step $\Delta t$ and add the energy as well as momentum source terms 
to the gas. 
\item Continue the above steps until the end of the simulation.
\end{enumerate}

\begin{figure}[htp]
\centering
\includegraphics[width=0.48\hsize]{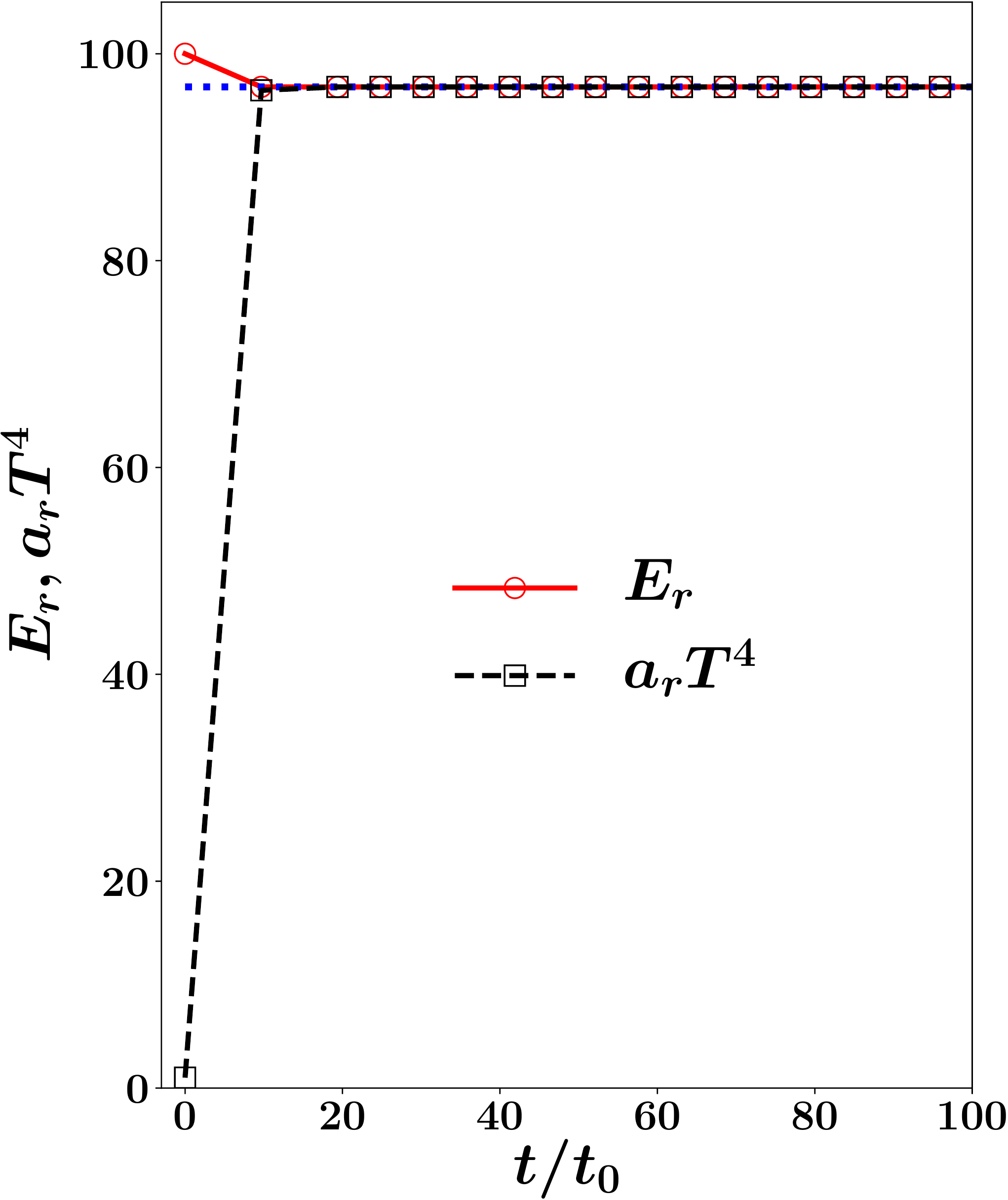}
\includegraphics[width=0.48\hsize]{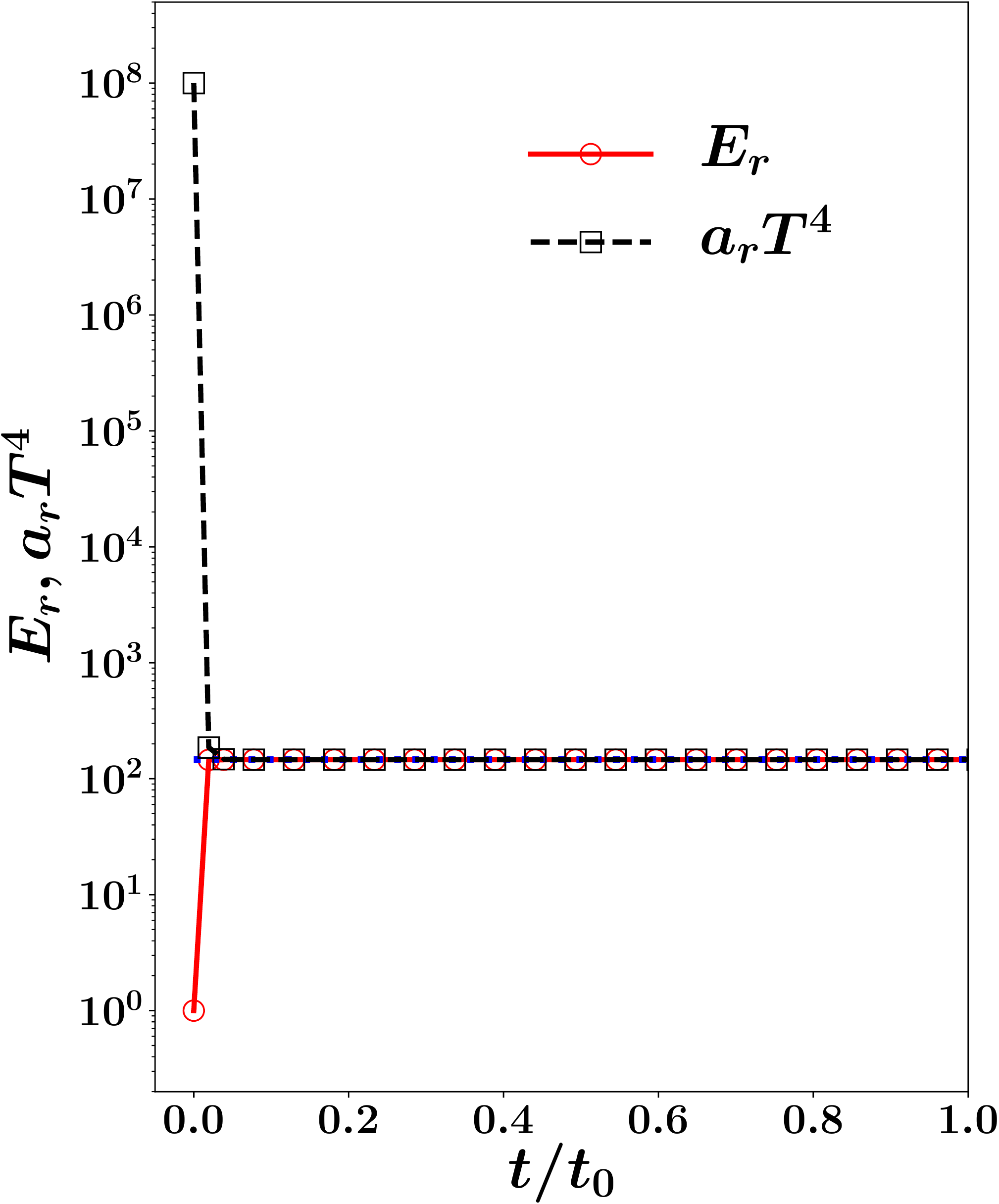}
\caption{Histories of $E_r$ and $a_rT^4$ when gas and radiation relax to the thermal equilibrium state from 
static and uniform initial conditions. The left and right panels show cases with different relative values between $a_rT^4$ and $E_r$. 
The time unit is $t_0=1/(c\rho\kappa_a)$ in the two plots. The dotted blue lines indicate the analytic solution 
during the thermal equilibrium state. }
\label{fig:thermal_equilibrium}
\end{figure}

\section{Tests of the Algorithm}
\label{sec:test}
The algorithm we developed here can in principle work in a wide range of parameter space in terms of optical depth and 
ratio between radiation pressure and gas pressure, since no approximation is made when we solve the transport equation. 
When the time step is made sufficiently small, the algorithm can also be reduced to the explicit scheme as described by \cite{Jiangetal2014}. 
However, in practice, convergent properties for our iterative scheme can be very problem dependent. 
We will repeat most of the tests done by \cite{Jiangetal2012} and \cite{Jiangetal2014} to explore the accuracy and efficiency of our algorithm. 
Unless otherwise specified, we will assume the temperature, density and length units to be $T_0,\rho_0,L_0$ respectively. The velocity unit $v_0$
 is chosen to be the isothermal sound speed corresponding to the temperature $T_0$ and the time unit is given by $L_0/v_0$. Radiation energy 
 density is scaled to $a_rT_0^4$ and radiation flux is scaled to $ca_rT_0^4$.  
 Solutions given in the dimensionless format can be converted to any 
 unit system given  the two dimensionless 
parameters $\Crat\equiv c/v_0$ and $\Prat=a_rT_0^4/\left(\rho_0 R_{\text{ideal}}T_0\right)$ \citep{Jiangetal2012}.

\begin{figure}[htp]
\centering
\includegraphics[width=1\hsize]{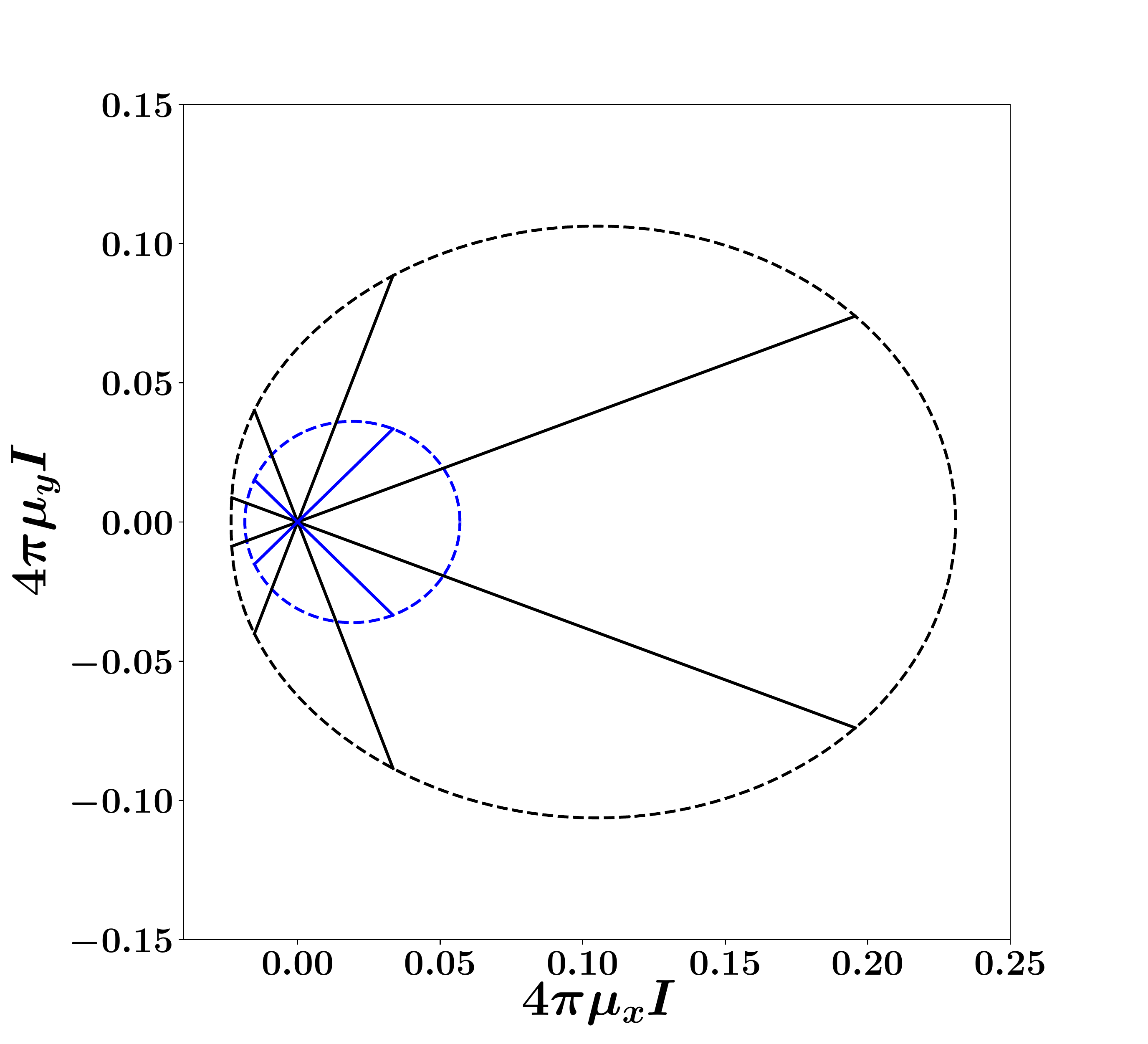}
\caption{Angular distribution of the lab frame specific intensities in the steady state for a moving uniform gas as described in Section \ref{sec:uniform}. 
The black and blue lines connect the origin and the points $(4\pi\mu_x I,4\pi\mu_y I)$ for the twelve angles we have in the numerical solution. 
All the angles represented by the blue lines 
have $\mu_z=0.333$ while the other angles represented by the black lines have $\mu_z=0.882$. 
The dashed blue and black circles represent analytical values for all the angles $(\mu_x,\mu_y)$ 
with the same corresponding $\mu_z$ respectively.  }
\label{fig:intensitymap}
\end{figure}

\begin{figure}[htp]
\centering
\includegraphics[width=1\hsize]{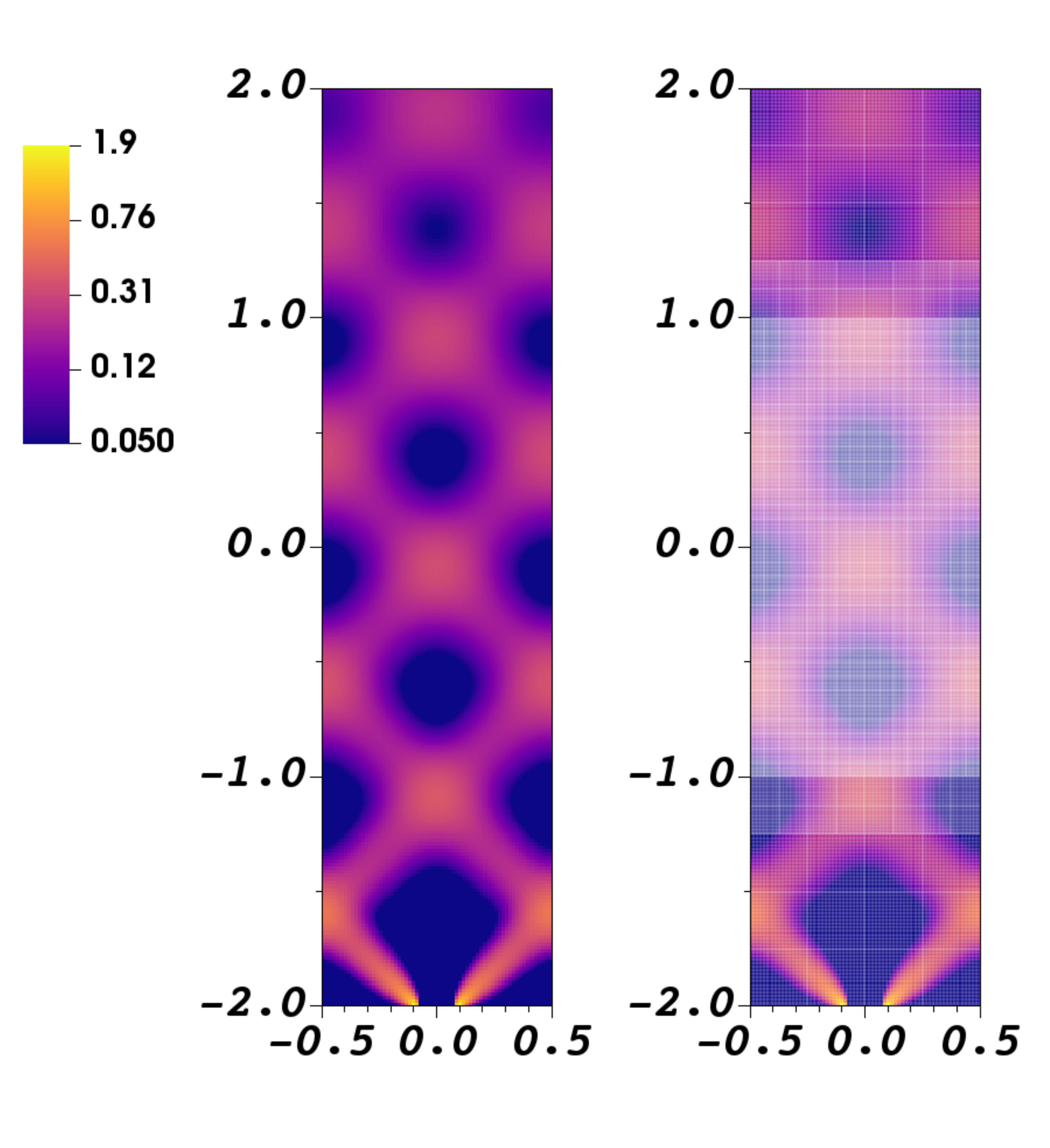}
\caption{Spatial distribution of mean radiation energy density $E_r$ after two beams of  photons propagate through the 
2D box with zero opacity. The two beams are injected from $x=\pm 0.1,y=-2$ with angles $\mu_x=\pm0.577,\mu_y=0.577$. 
The left and right boundaries are periodic while the two beams leave the box from the top. The left panel is the case with uniform resolution while 
static mesh refinement is used in the right panel as indicated by the overplotted grid lines. 
}
\label{fig:beam}
\end{figure}

\subsection{Thermal Equilibrium in a uniform medium}
\label{sec:uniform}
To test our treatment of the source term, we setup a uniform box in 2D covering the region $[0,1]\times [0,1]$. The box is initialized with 
uniform density $\rho=1$ and gas temperature $T=1$. The radiation field is initialized isotropically with the mean energy density $E_r=100$. 
We fix the two dimensionless parameters $\Prat$ and $\Crat$ to be $1$ and $100$ respectively. We use $32$ grid points for both direction. 
A constant absorption opacity $\rho\kappa_a=100$ is adopted for the test. The typical thermalization time scale is then $1/\left(\Crat\rho\kappa_a\right)=10^{-4}$ 
while the time step we use is roughly $0.001$. Therefore, in one time step, the system reaches the correct thermal equilibrium state and stays there as shown in the 
left panel of Figure \ref{fig:thermal_equilibrium}. The equilibrium state also agrees with the analytical solution very well as constrained by the total energy 
conservation. During the thermalization process, the algorithm can also keep the sign of $E_r-a_rT^4$ so that there is no oscillatory behavior as a function of time, which can 
happen for high order time integrators when the time step is much larger than the thermalization time scale. As another test, we change the initial condition to be 
$T=100,~ E_r=1$ and $\rho\kappa_a=1$ so that the gas is much hotter than the equilibrium state in this case. We can still get the correct equilibrium state even though there is a huge diffrence between $a_rT^4$ and $E_r$ initially as shown in the right panel of Figure \ref{fig:thermal_equilibrium}.

Our treatment of the source term is very similar to the scheme as described in \cite{Jiangetal2014}, where the flux divergence term is calculated explicitly. 
However, there are non-trivial differences for the term $\bn\cdot\bfnabla I$ even for the case with a uniform spatial distribution. When this term is calculated using  
the variables from the last time step, it is guaranteed to be zero to the level of round-off error as there is no spatial gradient at each step. This is not the case for the 
fully implicit scheme. During each cycle of the iteration when the gradient is calculated for each cell, variables in the neighboring cells are taken from last iterative step while we use the values from this step for the current cell. Before the system reaches the equilibrium state, the flux divergence term is actually not zero. After the solution converges, we recover the same solution as given by the explicit scheme with the accuracy set by the tolerance level. The Jacobi like iterative scheme we adopt also guarantees that all the cells are treated in the same way during each iteration. Therefore, we do not generate any artificial spatial gradient during the iteration. 

We have also tested the velocity dependent terms by giving the gas a uniform horizontal velocity $v_x=3$. Specific intensities are initialized isotropically in the lab frame with the mean energy density $E_r=1$. The initial gas temperature and density are $T=1,\rho=1$ respectively. We choose $\rho\kappa_a=1,\Prat=1$ and $\Crat=100$ for this test and use 12 angles per cell. The system settles down to a steady state with gas velocity reduced to $v_x=2.956$ as a small fraction of gas momentum is transferred to photons. The horizontal lab frame radiation flux in the steady state $F_{r,x}$ is very close to $v_x(E_r+P_{r,xx})$, which is the value given by the first order expansion in terms of $v/c$ in radiation moment equations \citep{Lowrieetal1999} since $v_x/\Crat$ is only $3\%$ in this test. During this process, the change of kinetic energy is converted to radiation energy and lab frame $E_r=1.13$ in the steady state. The specific intensities can also be compared with the analytical values directly, which are shown in Figure \ref{fig:intensitymap}. In steady state, the radiation field is isotropic in the co-moving frame. Angular distribution of specific intensities in the lab frame is then determined by the Lorentz transformation (equation \ref{eq:lorentz_transform}). The numerical solution agrees with the analytical values very well. 

Notice that the radiation field is not isotropic in the lab frame and the Eddington tensor is $P_{r,xx}/E_r=0.417$.  This will be different from the Eddington tensor returned by the VET method for this case since the short characteristic algorithm typically neglects the velocity dependent terms and it will return an Eddington tensor $1/3\bI$ in the lab frame.

\subsection{Crossing Beams in Vacuum}
Our iterative scheme is typically easier to converge when the source term dominates over the transport term, because this means the diagonal coefficients in the resulting matrix of equation \ref{eqn:In_final} are larger than the off-diagonal one. The most difficult scenario is the transport of photons through a medium with zero opacity, where all the source terms are zero and we only iterative over the transport term. To show how the algorithm works in this case, we inject two beams of photons from the bottom of a box covering the region $[-0.5,0.5]\times [-2,2]$. At $x=\pm 0.1, y=-2$, the specific intensities with $\mu_x=\pm 0.577,\mu_y=0.577$ are set to be $0.8$ while specific intensities along other angles are $0$. All the intensities inside the box are also set to $0$ initially. The box is periodic along $x$ direction and outflow at the top. Opacity coefficients of the gas are set to $0$ and the gas does not evolve in this test. The spatial resolution is $64\times 256$ for the horizontal and vertical directions and we use $4$ angles per cell. The time step is $4\times 10^{-3}$, which is also the light cross time across the vertical direction of the box. During the first step, it takes $100$ iterations to reach the relative error $10^{-3}$, which declines very slowly with more iterations. After one step, the two beams have crossed the whole simulation box. In about $10$ time steps, the system reaches steady state and the spatial distribution of $E_r$ is shown in the left panel of Figure \ref{fig:beam}. Compared with the same test done by \cite{Jiangetal2014}, the solution is very similar to the case with first order spatial reconstruction but more diffusive compared with the solution using second order spatial reconstruction. This is expected as our implicit algorithm only uses first order spatial reconstruction. The overall cost of the implicit algorithm is also comparable to the cost if the transport equation is solved explicitly for this case, because the number of iterations we need for convergence is now comparable to the ratio between speed of light and sound speed. Another reason that this test requires a lot of iterations for convergence is that the initial condition is far away from the steady state solution. The injected photons at the bottom boundary need to be communicated with the whole simulation box during the iterations. However, multigrid technique can significantly speed up the convergence in this case because the information at the boundary can propagation to the domain much faster at lower resolution levels.

Our algorithm is a finite volume scheme, which evolves the specific intensities using fluxes at cell faces as other cell centered variables such as density in {\sf Athena++}. Therefore, the existing mesh refinement module in the code \citep{Stoneetal2020} can be directly used for our RT algorithm. One example is shown in the right panel of Figure \ref{fig:beam}. The setup is the same as the left panel except that we refine the region $[-0.5,0.5]\times[-1,1]$ with two levels of refinements. Photons can cross the refinement boundaries smoothly without any numerical artifact.

\subsection{Shadow Test}
\label{sec:shadow}
One widely used test to demonstrate the difference between solving the angular resolved RT and radiation moment equations based on simplified assumptions of the closure relation is to capture the shadow cast by an optically thick cloud with beams of photons\citep{HayesNorman2003,Gonzalezetal2007}. Diffusion like approximation will always send photons to the region where radiation energy density gradient exists and thus cannot keep the shadow. Other approximations that treat the radiation field as a fluid (such as M1, \citealt{Gonzalezetal2007,McKinneyetal2013}) have their own caveats such as artifically merging photons coming from different directions. Our implicit scheme should get the correct shadows as we still solve the specific intensities along different directions independently. 

\begin{figure}[htp]
\centering
\includegraphics[width=1\hsize]{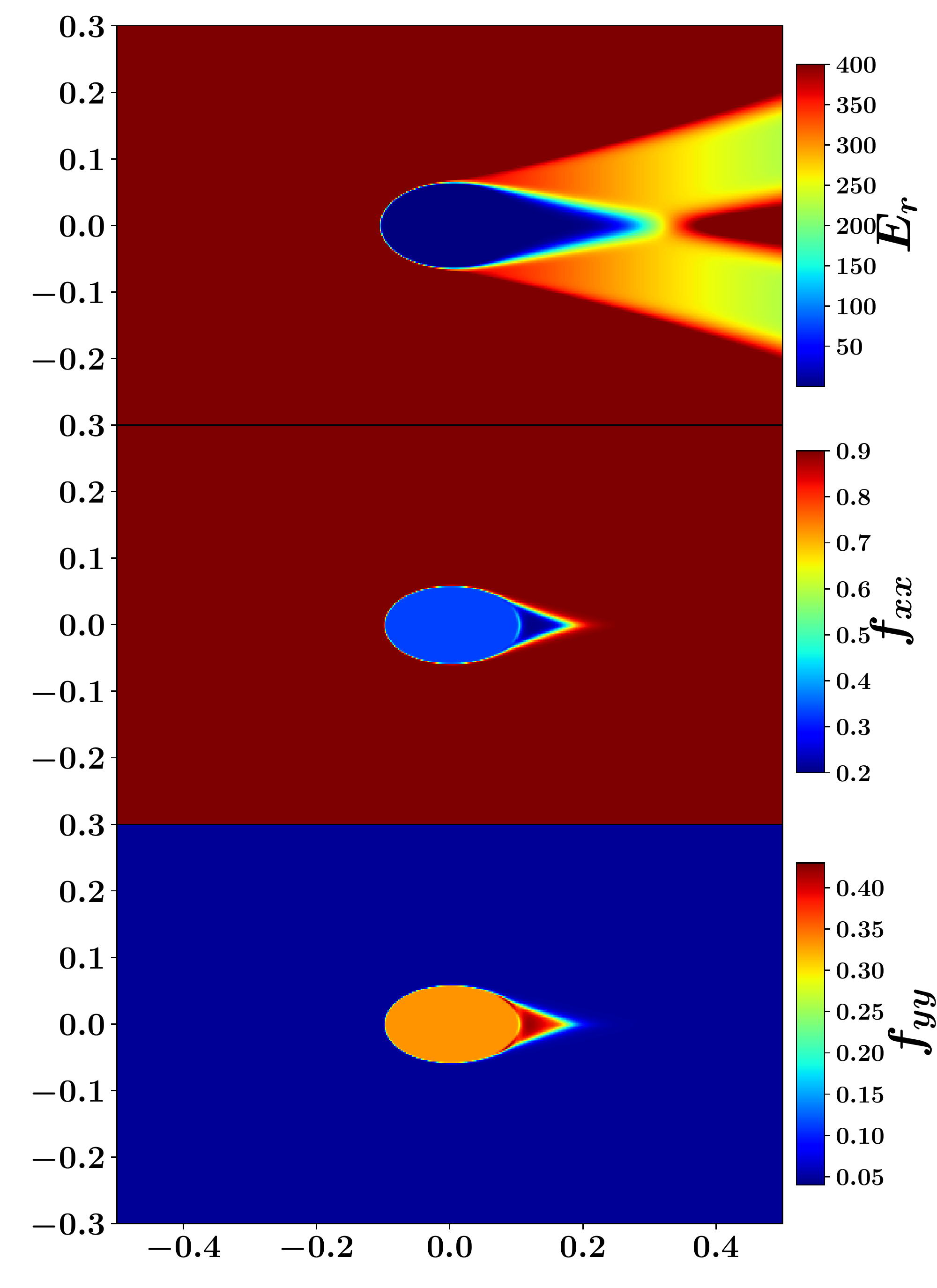}
\caption{The top panel shows the distribution of radiation energy density $E_r$ after two beams of photons pass an optically thick 
cloud, which is located at the center of the box as described in section \ref{sec:shadow}. The middle and bottom panels show the $x-x$ 
and $y-y$ components of Eddington tensor $f_{xx}$ and $f_{yy}$ respectively. }
\label{fig:shadow_static}
\end{figure}

\begin{figure}[htp]
\centering
\includegraphics[width=1\hsize]{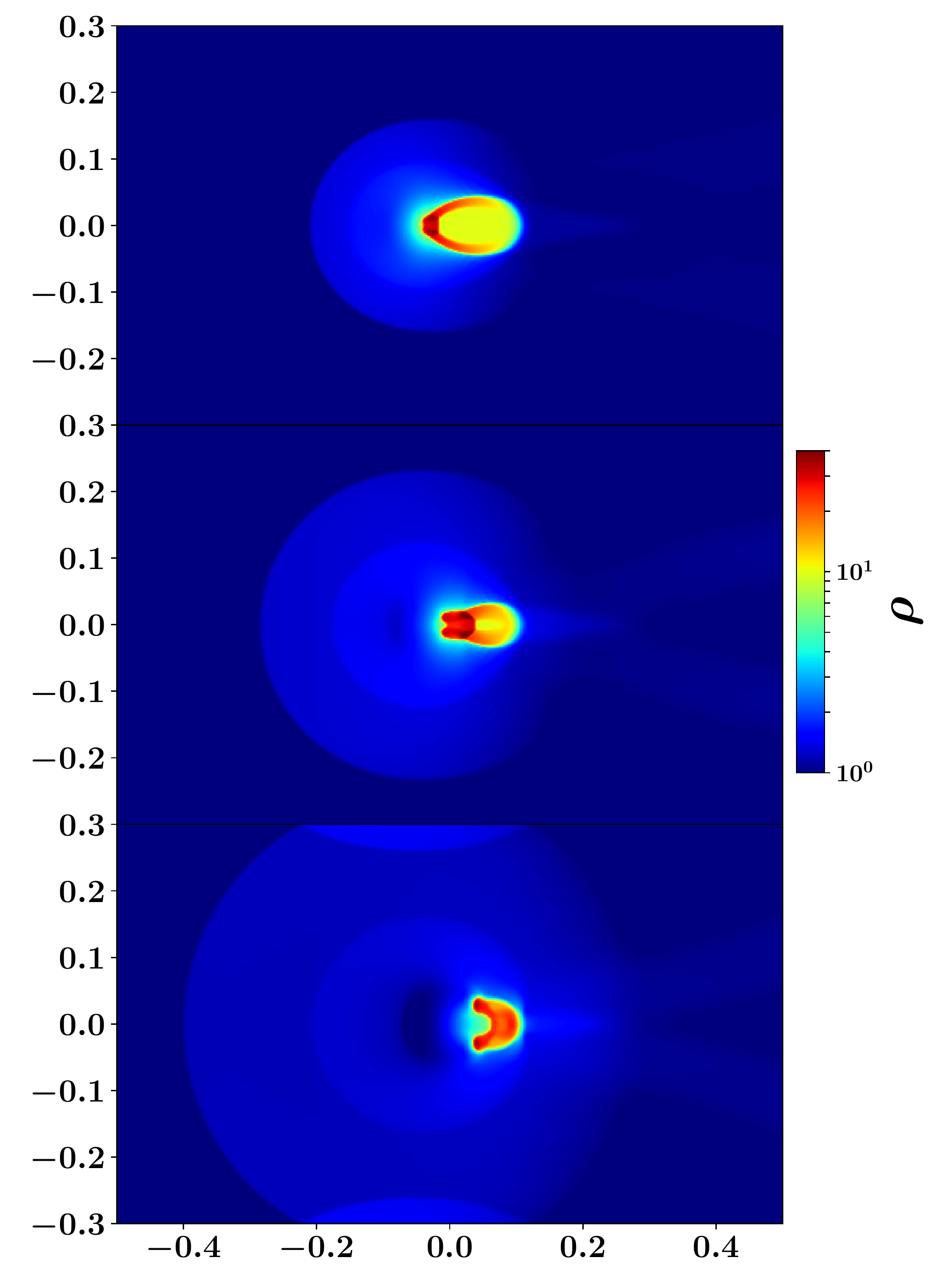}
\caption{Density evolution of an optically thick cloud irradiated by two beams of photons from the left boundary.  The three snapshots are shown 
at times $0.06t_0,0.1t_0,0.16t_0$ from top to bottom respectively, where $t_0$ is the sound crossing time across the whole box horizontally in the low density gas.}
\label{fig:shadow_dynamic}
\end{figure}

We used the same setup as in \cite{Jiangetal2012} and \cite{Jiangetal2014} for direct comparison. The simulation domain covers the cartesian box $[-0.5,0.5]\times [-0.3,0.3]$ with resolution $512\times 256$. We use 40 angles per cell for the 2D simulation. The gas density has the distribution $\rho(x,y)=1+9/\left[1+\exp\left[10\left((x/0.1)^2+(y/0.06)^2-1\right)\right]\right]$ while the temperature $T$ is set to achieve a constant gas pressure $P=1$ for the whole box. We use pure absorption opacity $\kappa_a=\rho T^{-3.5}$ so that the total optical depth across the box in the hot gas around the cloud is only $1$, while optical depth across the cloud is more than $10^5$. For the left boundary, specific intensities along the directions $\pm 15^{\circ}$ with respect to the horizontal axis have the fixed value $103.13$ but all the others are set $0$. For the right boundary, outgoing specific intensities are just copied from the last activate zones to the ghost zones while incoming specific intensities are $0$. Periodic boundary condition is used for the up and bottom boundaries.  We choose the dimensionless speed of light to be $\Crat=1.9\times 10^5$, which does not affect the steady state solution. All specific intensities are initialized to be $0$ and we keep the gas quantities fixed for the shadow test. It takes about $1000$ iterations to reach a relative error of $10^{-6}$ with three time steps and the radiation quantities are very close to the steady state values across the whole simulation box. The spatial distributions of mean radiation energy density $E_r$ and components of the Eddington tensor $f_{xx}$ and $f_{yy}$ are shown in Figure \ref{fig:shadow_static}. Our implicit algorithm gets the umbra and penumbra correctly and the results are very similar to the solutions returned by the explicit scheme as shown in \cite{Jiangetal2014} as well as the VET  method as described by \cite{Jiangetal2012}. 

One great advantage of our implicit algorithm is that it can also capture the time dependent evolution of the radiation field efficiently even though the speed of light is $1.9\times 10^5$ times the typical sound speed. To demonstrate that our algorithm can also capture the full dynamical evolution of the cloud,  we use the same setup as in the shadow test but evolve the hydrodynamic quantities with the radiation field together. We choose the temperature scale such that $\Prat=10^{-7}$ so that the radiation pressure of the injected photons from the left boundary is about $10^{-3}$ of the initial gas pressure. Notice that the static shadow test is independent of this ratio. The parameter is also chosen to match the same experiment done by 
\cite{Jiangetal2012}. Outflow boundary condition is used for the hydrodynamic variables at the left and right boundaries while periodic boundary is used for the top and bottom. The typical sound crossing time scale across the box is $t_0=1$ while the light crossing time in the low density gas is $\approx 5.2\times 10^{-6}t_0$. The photon diffusion time scale across the cloud is comparable to $t_0$. Since the effective temperature of the incoming radiation field is larger than the temperature of the cold gas by a factor of $30$, the gas will be heated up in the time scale of $2\times 10^{-5}t_0$. Density distributions of the cloud at three different times $t=0.06t_0, 0.1t_0,0.16t_0$ are shown in Figure \ref{fig:shadow_dynamic}. The cloud is heated up at the surface of the left side initially. The heated gas drives shocks into the cloud and some gas is also evaporated simultaneously. The shock propagates inside the cloud and it gets destructed eventually. The overall evolution is very similar to the results shown by \cite{Jiangetal2012} and \cite{Progaetal2014}, despite some small differences of the detailed structures. Compared with the VET scheme, there is a subtle difference on effectively what equations are solved. The VET scheme typically calculates the Eddington tensor at the \emph{ beginning} of each time step, which is then held fixed when the radiation moment equations and hydrodynamic equations are evolved for one time step. The algorithm developed here solves the specific intensities implicitly with the same time evolution as the hydrodynamic variables. If we integrate the specific intensities over the angular space, effectively we evolve the radiation moments using the Eddington tensor determined at the \emph{end} of each time step consistent with the implicit update of the radiation field. This will not make any difference for steady state solutions or if the time step is sufficiently small. However, when the time step is much larger than the light crossing time for each cell, change of the Eddington tensor due to time evolution of the radiation field within one time step may not be negligible. Future studies are needed to quantify the difference. 

\begin{figure}[htp]
\centering
\includegraphics[width=1\hsize]{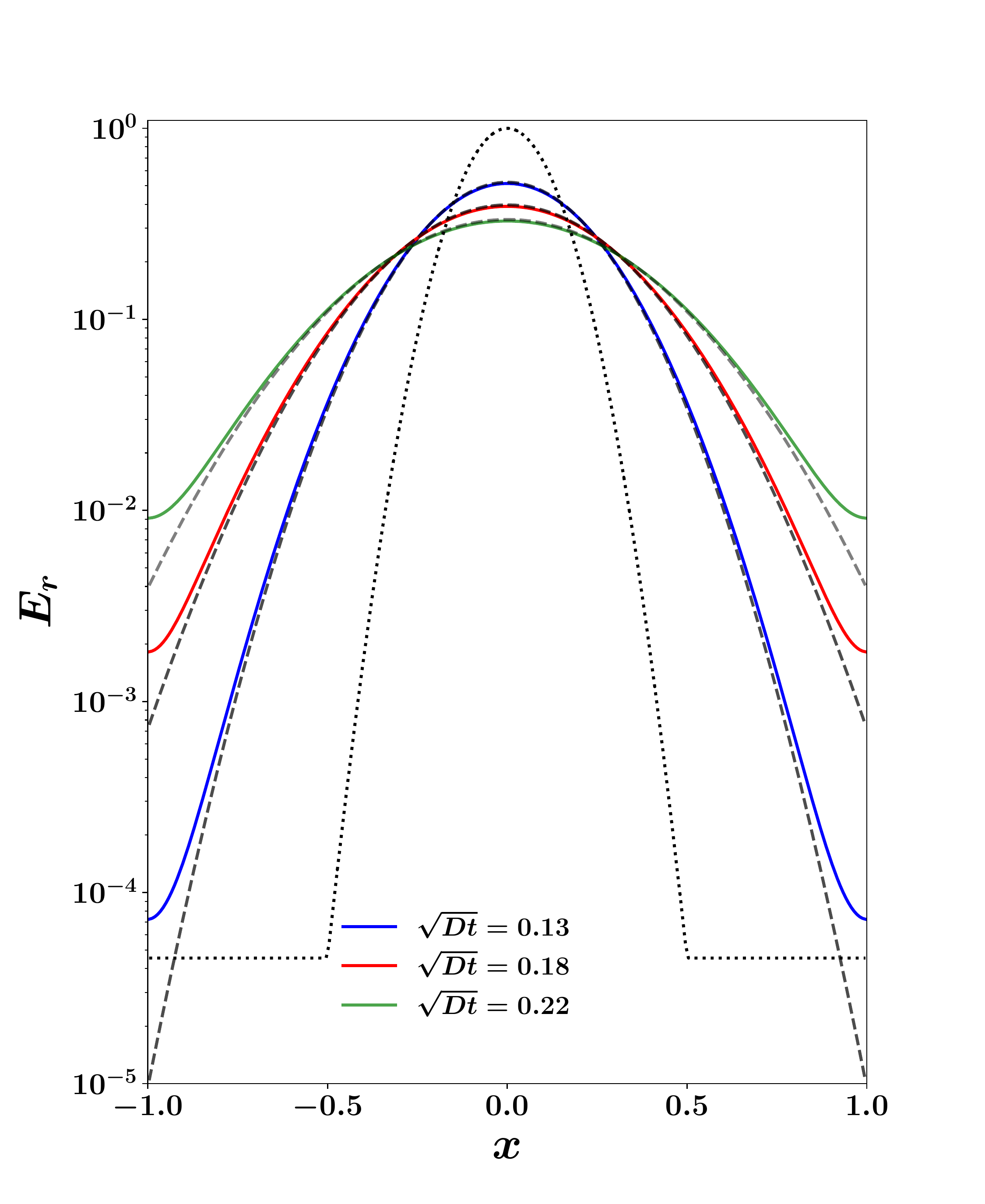}
\caption{Profiles of radiation energy density $E_r$ at three different times corresponding to $\sqrt{Dt}=0.13, 0.18, 0.22$ from an initial Gaussian profile as indicated by 
the dotted black line. The diffusion coefficient is $D=\Crat/\left(3\rho\kappa_s\right)=8.33\times 10^{-5}$ for this test. The solid lines are numerical solutions while the dashed lines are analytical solutions to the diffusion equation. }
\label{fig:static_diffusion}
\end{figure}

\begin{figure}[htp]
\centering
\includegraphics[width=1\hsize]{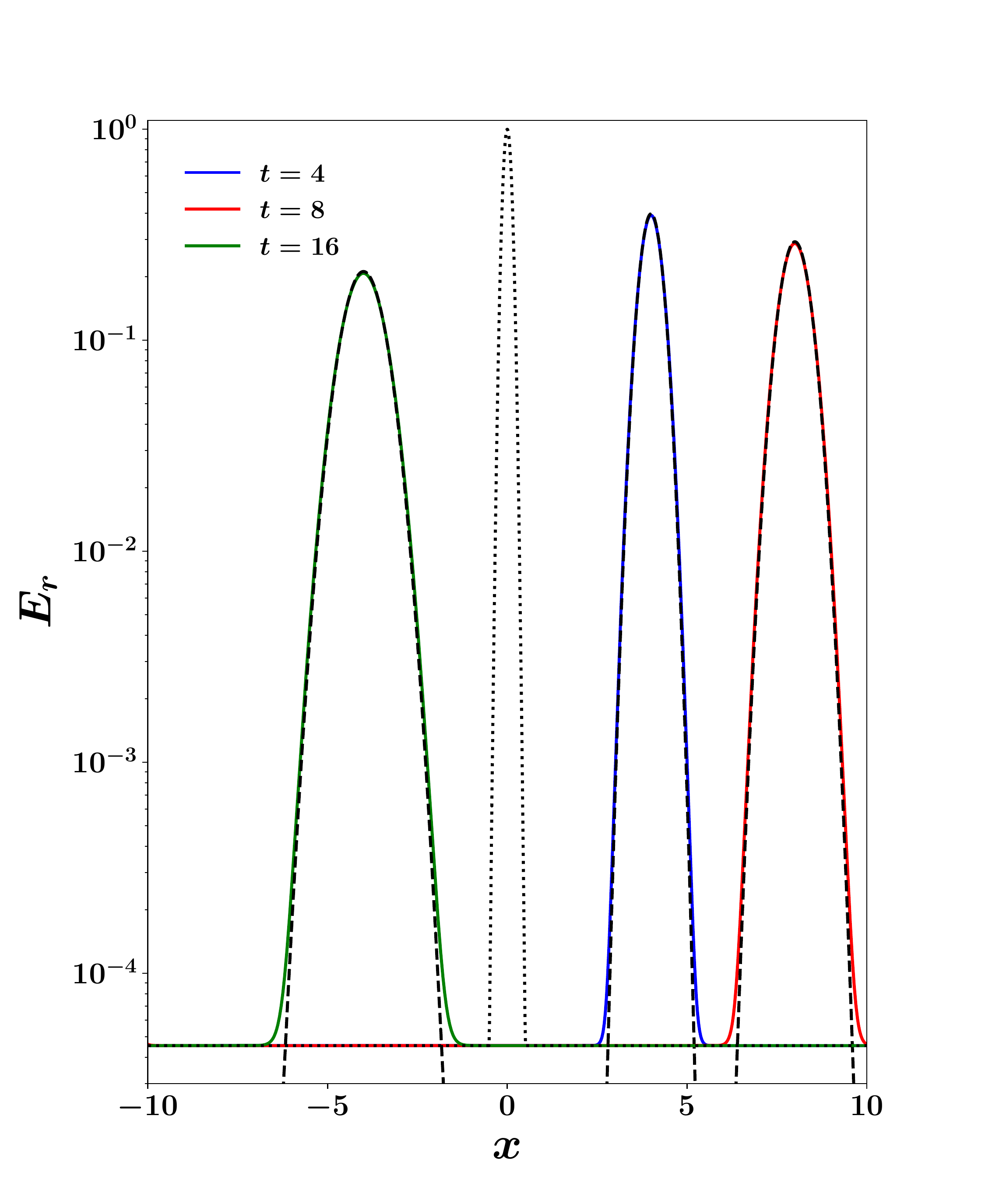}
\caption{Diffusion of photons in a moving background as described in section \ref{sec:diff}. The gas has a constant scattering opacity $\rho\kappa_s=4\times 10^4$, 
velocity $v=1$ while the speed of light is $10^3$ in this test. The colored solid lines are numerical solutions at 
times $t=4,8,16$ respectively, while the corresponding analytical solutions are indicated by the black dashed lines. The dotted line at $x=0$ is the initial profile. }
\label{fig:dynamic_diffusion}
\end{figure}

\subsection{Static and Dynamic Diffusion}
\label{sec:diff}
Photon diffusion in the optically thick regime, which is easy to solve with the classical diffusion approximation by design, turns out to require careful treatment when the 
full time dependent transport equation is solved. This is because in the very optically thick regime, the numerical diffusion associated with the standard HLLE solver can be much larger than the physical diffusive flux (see the Appendix of \citealt{Jiangetal2013b}). To demonstrate that our treatment of the transport term as described in Section \ref{sec:transport} is sufficient to capture the photon diffusion correctly, we set up a radiation field with mean energy density $E_r(x)=\exp\left(-40x^2\right)$ in the box $x\in\left(-0.5,0.5\right)$. For $|x| > 0.5$, we set $E_r=\exp\left(-10\right)$. We use $256$ grid points for the whole simulation box $(-1,1)$ and 2 angles per cell so that the Eddington tensor is always $1/3$. Outflow boundary condition is used for this test. The gas has a constant scattering opacity $\rho\kappa_s=4\times 10^4$ and we do not evolve the gas quantities for this test. We choose the dimensionless speed of light to be $\Crat=10$. 
Evolution of $E_r$ can be described by the solution to the diffusion equation as \citep{SekoraStone2010,Jiangetal2014}
\begin{eqnarray}
E_r(x,t)=\frac{1}{(160 Dt+1)^{1/2}}\exp\left(\frac{-40x^2}{160 Dt+1}\right),
\label{eq:sol_diff}
\end{eqnarray}
where $D\equiv \Crat/(3\rho\kappa_s)$ is the diffusion coefficient. The specific intensities are initialized isotropically with the mean energy density matching the initial profile of $E_r$. Profiles of $E_r$ at three different times from the numerical solution are shown in Figure \ref{fig:static_diffusion}, which match the analytical solution very
well. Notice that because we do not set the boundary condition to match the diffusion solution exactly, we can only compare the region roughly $\sqrt{Dt}$ away from the boundary. We have also tried cases with larger opacity and our algorithm can still get the solution correctly. However, we need to use a larger value of the parameter $\tau_c$ (equation \ref{eq:taucell}) to get the same accuracy with a larger $\kappa_s$.

Another important regime is dynamic diffusion, where the photons are advected with the flow while diffusion happens. We use a similar setup as for the static diffusion case but extend the simulation domain to $(-10,10)$ with 1280 grid points. Periodic boundary condition is used so that photons are not advected out of the box. We give the flow a constant velocity $v=1$. The dimensionless speed of light is also increased to $\Crat=1000$ for this test so that the typical diffusion speed is not negligible compared with the advection speed. Profiles of the mean radiation energy densities at time $t=4,8,16$ are shown as the solid lines in Figure \ref{fig:dynamic_diffusion}. The analytical solutions can also be estimated by replacing the position $x$ in equation \ref{eq:sol_diff} with $x-vt$ to the first order of $v/c$, which is $10^{-3}$ in this test.  The analytical solutions are shown as the black dashed lines in Figure \ref{fig:dynamic_diffusion} and they match the numerical solution very well.

\begin{figure}[htp]
\centering
\includegraphics[width=1\hsize]{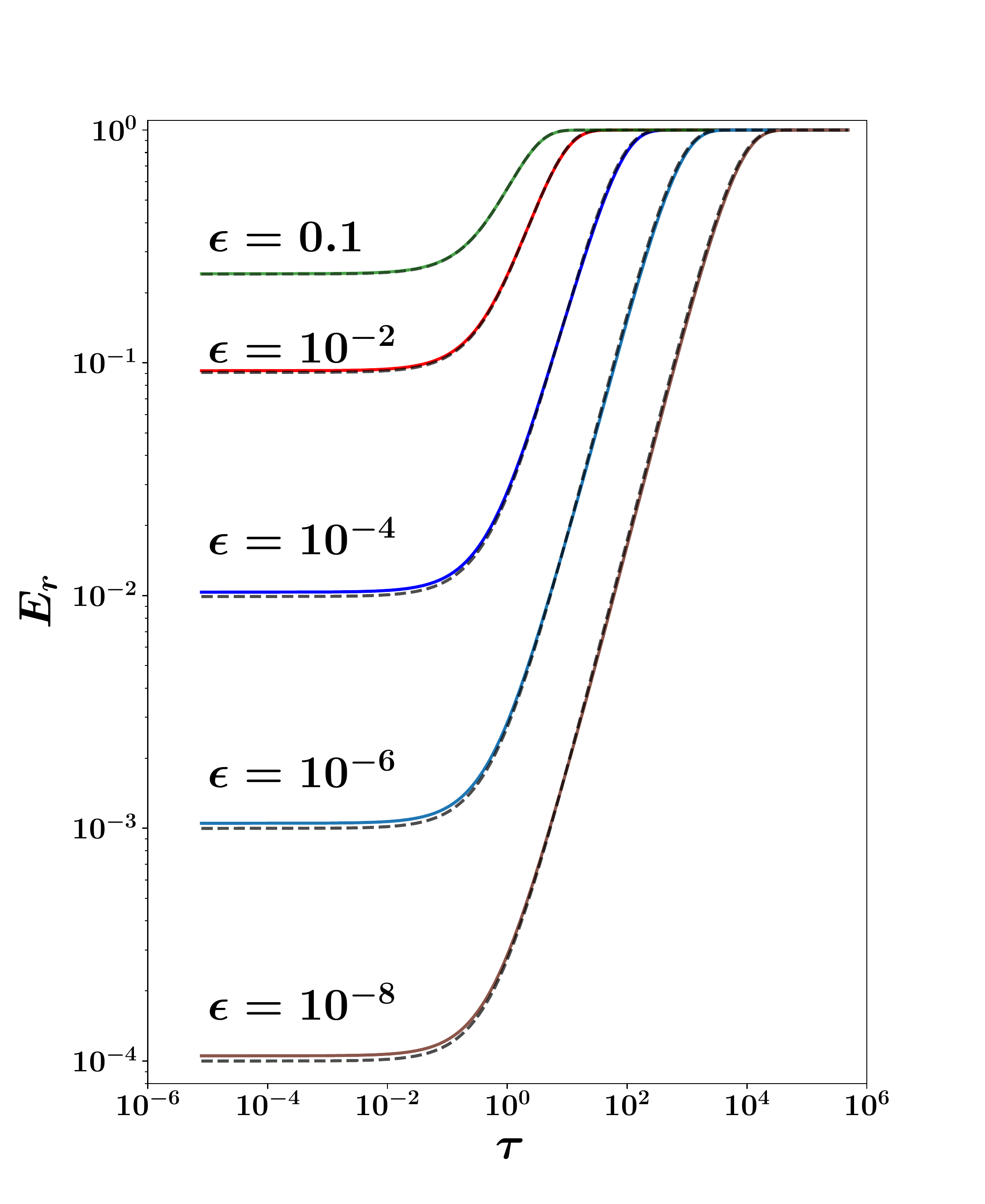}
\caption{Profiles of radiation energy density as a function of optical depth for the atmosphere test as described in Section  \ref{sec:atmosphere}. 
The solid lines are numerical solutions while the dashed black lines are corresponding analytical solutions. The ratio between absorption  and total 
opacity $\epsilon$ varies from $0.1$ to $10^{-8}$ as indicated for each group of lines in the plot. }
\label{fig:atmosphere}
\end{figure}

\begin{figure}[htp]
\centering
\includegraphics[width=1\hsize]{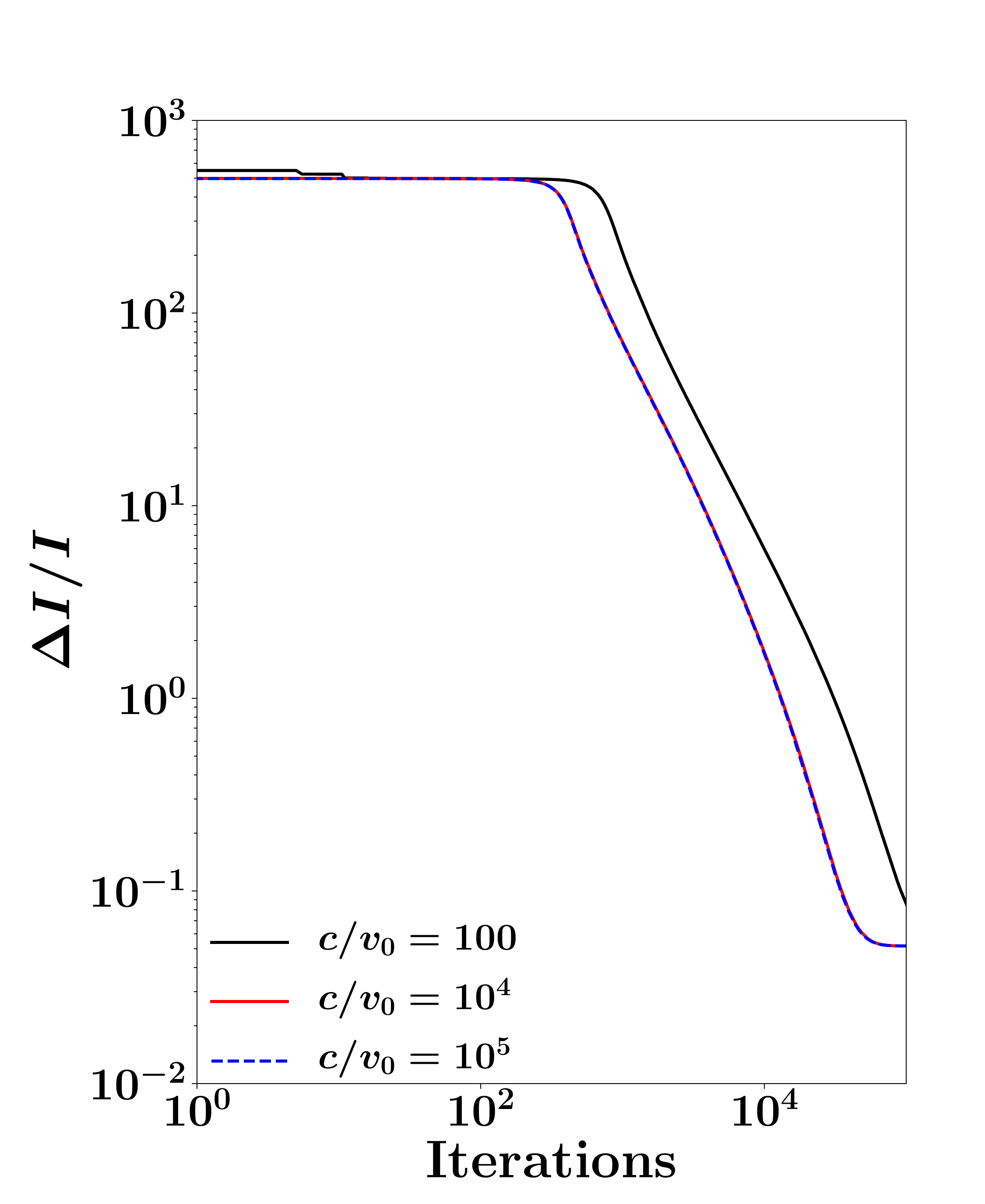}
\caption{Change of the relative error for specific intensities at $x=10$ of the atmosphere test described in Section \ref{sec:atmosphere} as a function of the total number of iterations. 
This is measured for the case with destruction fraction $\epsilon=10^{-6}$. Different lines represent cases with three different values of speed of light we used compared with the fiducial sound speed as indicated in the legend. 
}
\label{fig:converge}
\end{figure}

\subsection{Non-LTE Atmosphere}
\label{sec:atmosphere}
The test of a uniform temperature, scattering dominated atmosphere done by \cite{Davisetal2012} is useful to show how our iterative scheme 
performs when there is a large dynamical range of radiation energy density with both optically thin and optically thick regions. Particularly, we can compare 
the performance of our iterative scheme with the classical short characteristic method for the problem that the classical method is efficient to solve. 
We use $1280$ grid points covering a simulation domain $x\in [-10,10]$ in 1D. Gas density follows the exponential profile $\rho=10^{-3}\exp(10-x)$ and gas 
temperature is fixed to be $T_i=1$. The ratio between absorption and total opacity is $\epsilon$, which is taken to be a constant in the whole simulation domain.  Therefore, 
the scattering and absorption opacity are  $\kappa_a=\epsilon$ and $\kappa_s=1-\epsilon$. The gas quantities are fixed to be the initial conditions with zero velocity and does not evolve. Since our algorithm always solves the gas internal energy and radiation field together, we simply reset the gas quantities to be the initial profiles at the end of each step. We use one angle per octant for this test so that the Eddington tensor is always $1/3$. At the bottom boundary, we set the specific intensities to be the isotropic value $a_rT_i^4/(4\pi)$. At the top boundary, the incoming specific intensity is set to be $0$ while the outgoing specific intensity is copied from the last active zone to the ghost zone. Specific intensities inside the simulation domain is initialized to be $a_rT_i^4/(4\pi)$.

The steady state profile of radiation energy density for this atmosphere setup is
\begin{equation}
E_r=a_rT_i^4\left[1-\frac{\exp\left(-\sqrt{3\epsilon}\tau\right)}{1+\sqrt{\epsilon}}\right],
\end{equation}
where $\tau\equiv \int_x^{10}\rho dx$ is the total optical depth integrated from $x=10$. Upward and downward specific intensities are given by 
$I_{+}=(1/4\pi)\left[E_r+(1/3\mu_x)\partial E_r/\partial \tau\right]$ and $I_{-}=(1/4\pi)\left[E_r-(1/3\mu_x)\partial E_r/\partial \tau\right]$ respectively.
The steady state numerical solutions for $\epsilon$ varying from $0.1$ to $10^{-8}$ 
are shown as the solid lines in Figure \ref{fig:atmosphere}, which agree with the analytical solutions very well. Our algorithm requires more iterations to reach the same level of accuracy when $\epsilon$ decreases. The relative error is also larger in the optically thin surface and the error increases with reducing $\epsilon$ for a fixed spatial resolution. All these properties are very similar to results given by the short characteristic method for this test problem \citep{Davisetal2012}. We have also tried the test in a 3D domain with the atmosphere profile aligned with the $x$ axis. We use periodic boundary condition for $y$ and $z$ directions in this case. The steady state solutions are identical to the 1D case and the rate of convergence is also the same. The overall cost is just proportional to the total number of grid points. 

To quantify the rate of convergence for our algorithm and compare with the short characteristic method, we pick the $\epsilon=10^{-6}$ case and measure the relative error $\Delta I/I$ of the specific intensities at $x=10$ after each iteration. Here we calculate $\Delta I$ as the difference between the solution at each iterative step and the analytical solution. We only focus on the location at $x=10$ because the difference between the initial condition and the analytical solution is the largest there. It is also the most optically thin region and takes more iterations to converge. The criterion is different from what \cite{Davisetal2012} used because we do not iterate over the source term. But both criteria serve the same purpose. Notice that the steady state solution does not depend on the speed of light. However, our algorithm always solves the time dependent transport equation and relaxes to the steady state solution with the time step $\Delta t$ fixed by the gas sound speed $v_0$. We have tried three different values of $c/v_0$ and the changes of $\Delta I/I$ with number of iterations are shown in Figure \ref{fig:converge}. When $c/v_0$ is small, although the implicit scheme is easier to converge per time step, it actually takes more total number of iterations to get the steady state solution. When $c/v_0$ is larger than $10^4$ so that $c\Delta t$ is larger than the length of the simulation box, the convergence rate is independent of $c/v_0$ anymore. 
Our implicit algorithm also takes more iterations to converge with this initial condition compared with the short characteristic method. This test demonstrates that the implicit scheme is not optimal just to find steady state solution as we are solving the full time dependent RT, although we can still get the correct answer.

\subsection{Homogeneous Sphere}
\label{sec:homo_sphere}
Solving the RT equations in 1D spherical polar coordinate can be useful for many systems that are close to spherical symmetric (such as stars), particularly with future development of multiple frequency groups. Spherical polar angular systems as described in section \ref{sec:sphere_angles} need to be used in this case because of the assumed symmetry. One useful test for this case is the homogeneous sphere problem, which is widely used to test the numerical scheme for radiation and neutrino transport \citep{Abdikamalovetal2012,Radiceetal2013}. For this test, we setup the gas with a constant density $\rho_0=1$ and temperature $T_0=1$ from radius 0.05 to $R_0=1$. Between radius 1 and the outer boundary, which is chosen to be 7, the density is $10^{-7}$ and temperature is $10^{-3}$. We assume absorption opacity $\rho_0\kappa_a=10$ for radius smaller than 1 and the opacity is 0 in other region. The gas is held with zero velocity and does not evolve.  The units can be chosen to match any system and do not affect the solution to the transport equation.  We set the specific intensity to be isotropic with the value $a_rT_0^4/(4\pi)$ at the inner boundary. For the outer boundary, we require the outgoing specific intensities to be continuous while the incoming specific intensities to be 0. We use 1000 uniformly distributed grid points for the spatial resolution and 40 angles per cell for the specific intensities. This setup covers both the optically thick and a sharp transition to the optically thin region, which can be challenging if the transport equation is not solved accurately. The specific intensities are initialized to be the isotropic value $a_rT_0^4/(4\pi)$ in the whole grid. We choose the speed of light to be 100 times the gas sound speed and the code finds the steady state solution in roughly 20 iterations. Notice that the steady state solution does not depend on the speed of light. 

This setup also has an analytical solution, which we can compare with. The specific intensities as a function of radius $r$ and angles $\bn$ are \citep{Smitetal1997}
\begin{align}
I(r,\bn)=\frac{a_rT_0^4}{4\pi }\left[1-\exp\left(\rho_0\kappa_a s(r,n_r)\right)\right],
\end{align}
where $n_r$ is the radial component of the unit vector $\bn$ and the function $s(r,n_r)$ is defined as
\begin{align}
s(r,n_r)=
\begin{cases}
      r n_r + R_0g(r,n_r) &  \text{if $r<R_0$}  \\
     2R_0g(r,n_r) & \text{if $r\geq R_0$  \&}  \\
      ~ & \text{$\sqrt{1-\left(R_0/r\right)^2}\leq n_r \leq 1$}   \\
       0 & \text{otherwise},
\end{cases}
\end{align}
 with the function $g(r,n_r)$ to be
 \begin{eqnarray}
 g(r,n_r)=\sqrt{1-\left(\frac{r}{R_0}\right)^2\left(1-n_r^2\right)}.
 \end{eqnarray}
Moments of the radiation field ($E_r, F_r, P_r$) can be calculated directly via angular quadrature of $I(r,\bn)$.
 
\begin{figure}[htp]
\centering
\includegraphics[width=1\hsize]{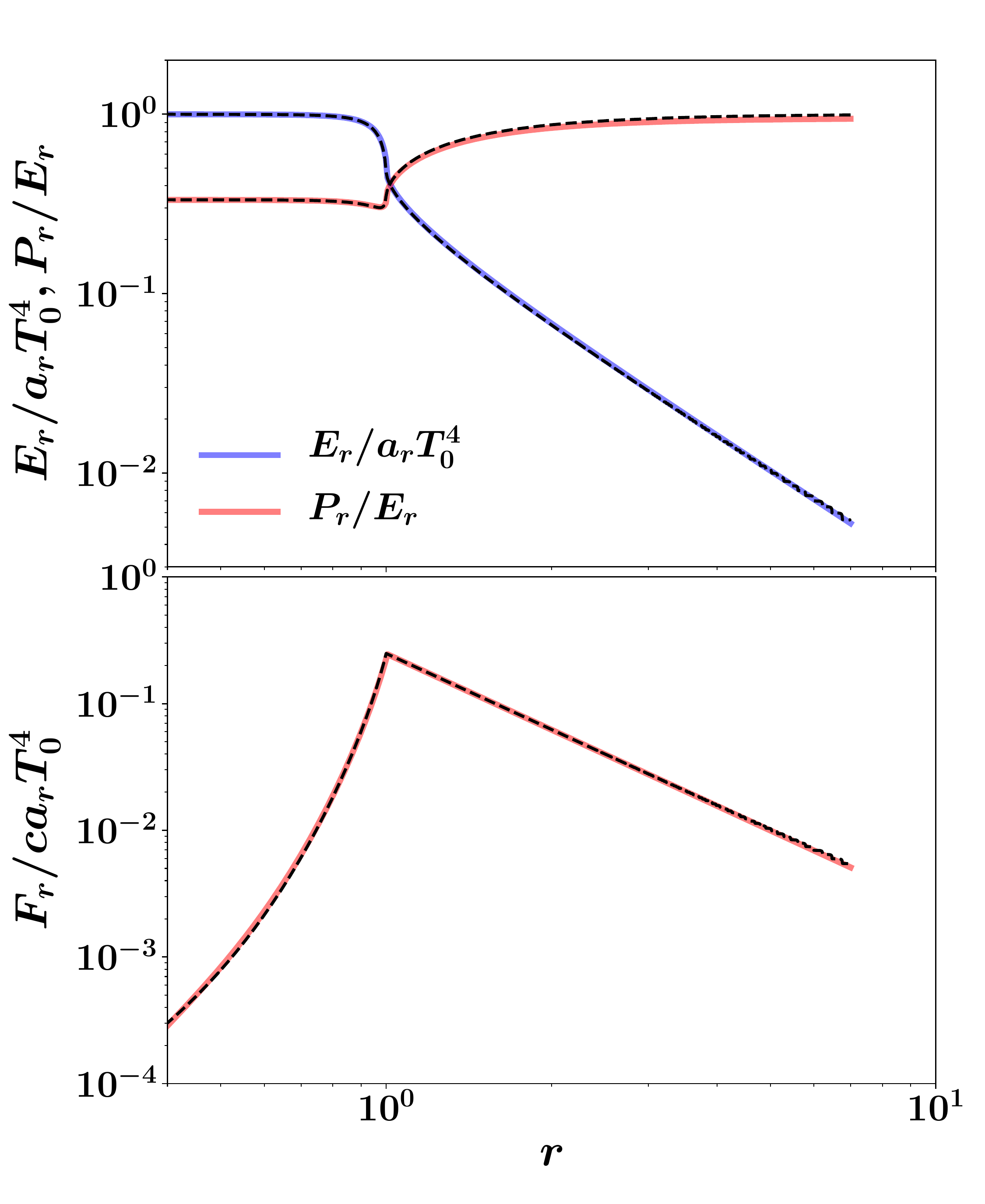}
\caption{Profiles of radiation energy density $E_r$, Eddington tensor $P_r/E_r$ (the top panel) and radiation flux $F_r$ (the bottom panel)
from the test of homogeneous sphere as described in section \ref{sec:homo_sphere}. The solid lines are from the numerical solution while the analytical solutions are given by 
the dashed lines.  }
\label{homo_sphere}
\end{figure}

Radial profiles of $E_r, F_r$ and $P_r/E_r$ from the numerical solution are shown as the solid lines in Figure \ref{homo_sphere}, 
which agree with the analytical solutions very well. Particularly, our numerical algorithm can capture the sharp transition at $r=1$ without showing any artificial numerical effect. 
The Eddington tensor $P_r/E_r$ is $1/3$ deep in the sphere. It is slightly smaller than $1/3$ around $r=1$. At large distance, the sphere becomes a point source effectively and only outgoing specific intensities pointing radially are non-zero. Therefore, the Eddington tensor approaches 1. The slight difference between the numerical and analytical solutions for $P_r/E_r$ is due to the finite angular resolution to resolve the angular domain $\sqrt{1-\left(R_0/r\right)^2}\leq n_r \leq 1$ at large $r$.

\begin{figure}[htp]
\centering
\includegraphics[width=1.0\hsize]{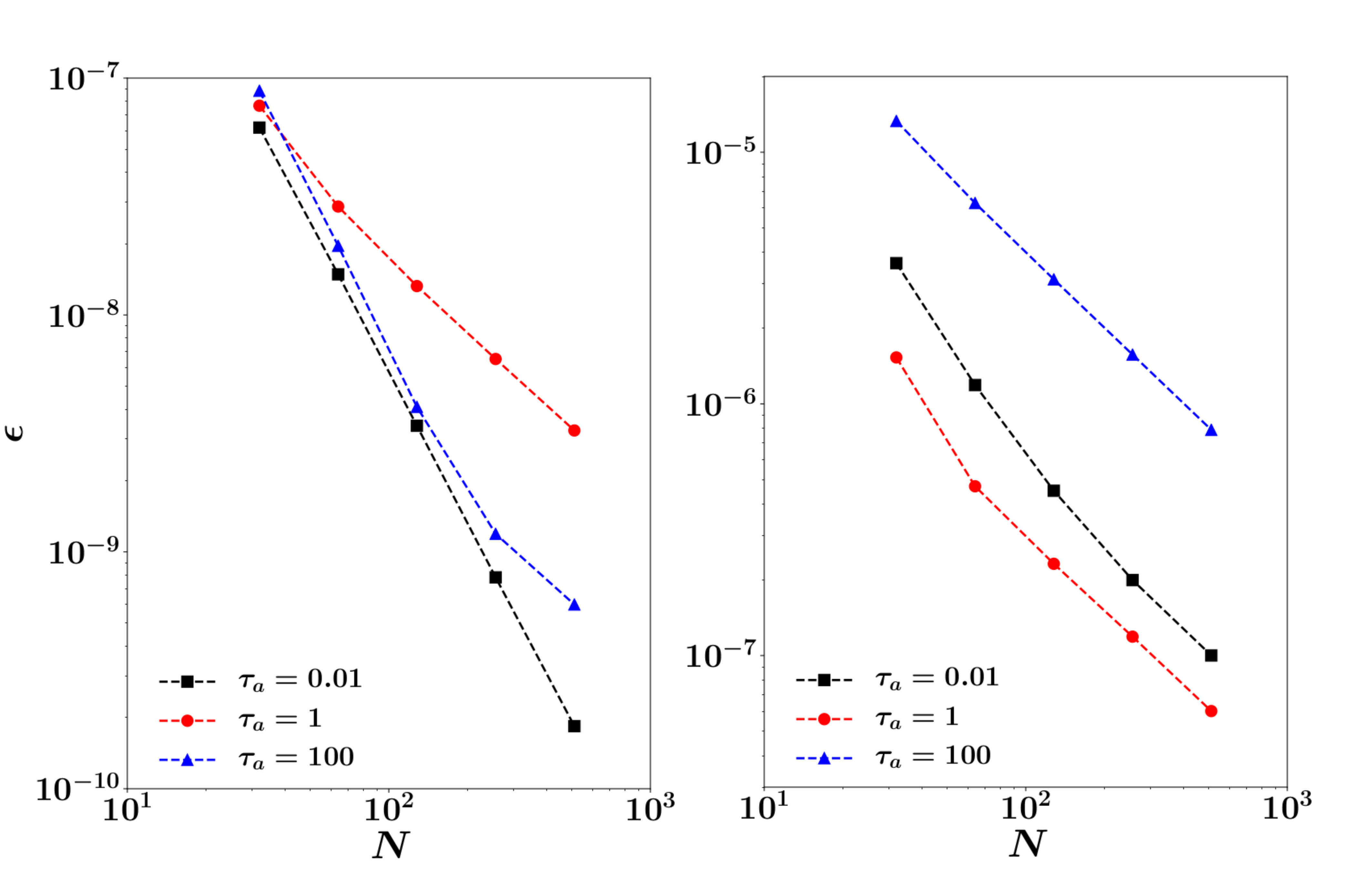}
\caption{Convergence of the L1 error $\epsilon$ for the linear wave tests as a function of spatial resolution $N$. The dots with different colors are 
for different optical depth per wavelength as labeled in the plot. The left and right panels are for background states with $\Prat=0.01$ and $10$ respectively. }
\label{linearwave_conv}
\end{figure}

\begin{figure}[htp]
\centering
\includegraphics[width=1\hsize]{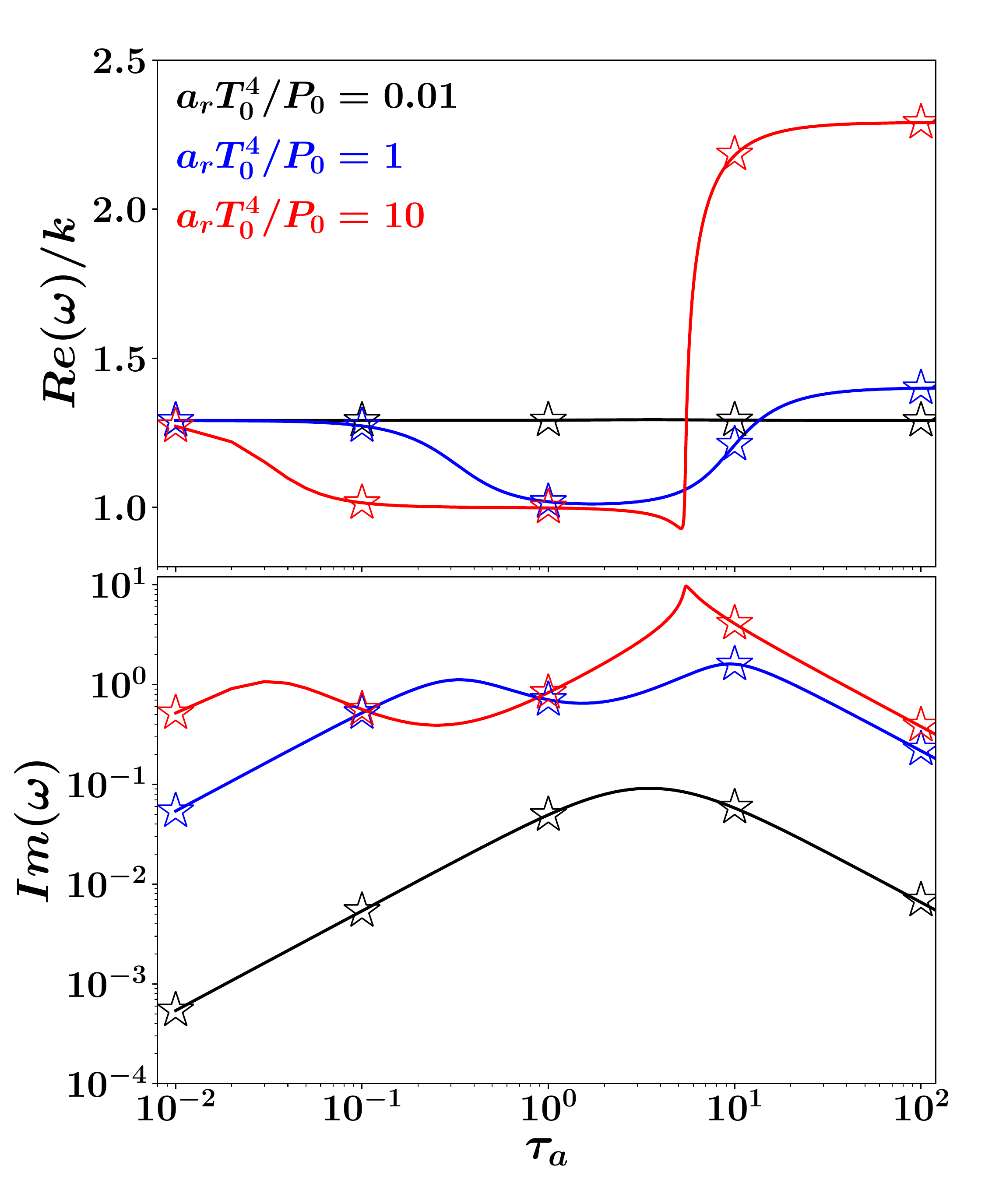}
\caption{Comparison of the propagation speed (the top panel) and damping rate (the bottom panel) 
for the radiation modified linear waves between the numerical calculated values (the dots) and analytical solutions (the solid lines). 
The lines with black, blue and red colors are for background states with $\Prat=0.01,1$ and $10$ respectively.}
\label{linearwave_sol}
\end{figure}

\subsection{Convergence of Linear Waves}
\label{sec:linearwave}
Properties of convergence for the full radiation MHD algorithm can be checked with linear wave tests. The linearized equations we solve here are identical to what are evolved by \cite{Jiangetal2014}. In order to compare directly, we adopt the same setup here. The background gas has uniform density $\rho=1$, and temperature $T=1$ with adiabatic index $\gamma=5/3$. We use one angle per octant with mean radiation energy density $E_r=1$ so that the Eddington tensor is always $1/3\bI$ as used in the analytical solutions. We fix the dimensionless speed of light to be $\Crat=10$ and only use a constant pure absorption opacity.  We use a 2D simulation box with size $L_x=1, L_y=1$ and  change the spatial resolution from $32\times 32$ to $512\times 512$. Periodic boundary conditions are used for both directions. We measure the L1 error $\epsilon$ between the numerical solutions and the analytical solutions over the whole simulation box for each resolution and check how the errors change with resolution. We have also tried the same test in 3D and get the same results, although these 3D tests are more expensive to run. We initialize the calculation with eigenmodes of the radiation modified sound waves as given by the Appendix B of \cite{Jiangetal2012} and stop the calculation after the waves propagate one period along $x$ direction. 

The two key dimensionless parameters that control the properties of radiation modifieid sound waves are optical depth per wavelength length $\tau_a$ and the ratio between radiation pressure and gas pressure $\Prat$ in the background state after $\Crat$ is fixed. Figure \ref{linearwave_conv} shows the change of $\epsilon$ with resolution $N$ for three different values $\tau_a$ and two different values of $\Prat$. When the error is dominated by the hydrodynamic part as in the optically thin regime, small radiation pressure or low resolution, the error decreases with increasing resolution as $N^2$. When the error is dominated by the RT module, it only shows first order convergence as expected. Part of the truncation error in RT is due to different upwind directions for specific intensities propagating along opposite directions as explained in section \ref{sec:transport}. For the case with the highest resolution $N=512$, we find that increasing $\tau_c$ (equation \ref{eq:taucell}) in the HLLE flux by a factor of 2 from our default value can help decrease the L1 error. 

We can also measure the numerically calculated propagation speed and damping rate of the linear waves  with the resolution $N=256$, which are shown in Figure \ref{linearwave_sol} for $\tau_a$ varying from $10^{-2}$ to $10^2$ with three different values of $\Prat$ in each case. When the optical depth is small, our algorithm is able to capture both the adiabatic sound wave with $\Prat=0.01$ and isothermal sound wave with $\Prat=10$. When $\tau_a$ is larger than $10$ so that radiation and gas are tightly coupled, we are also able to capture the radiation driven acoustic modes correctly. The results are the same as shown in Figure 10 of \cite{Jiangetal2014} as the same parameters are used. But the time step in this algorithm is a factor of 10 larger compared with the case when the time step is limited by the speed of light.

\subsection{Radiation Shocks}
\label{sec:rad_shock}

Another challenging test for the full algorithm is the radiation modified shocks, which check the numerical scheme in the non-linear regime. Particularly, we can see how good the 
source terms balance the transport terms for both the RT and hydrodynamic part, which is crucial to get the structures of the shock correctly, 
even though the time step is much larger than the light crossing time per cell. Since our algorithm has the full angular distribution of the radiation field, we will compare with the 
semi-analytical solutions provided by \cite{Fergusonetal2017}, which gets the radiation shock solutions by solving the time independent RT and hydrodynamic equations numerically without 
making any assumption on the closure relation. These are the same radiation shock solutions that are used to test the explicit RT algorithm in \cite{Jiangetal2014}.

\begin{figure}[htp]
\centering
\includegraphics[width=1.0\hsize]{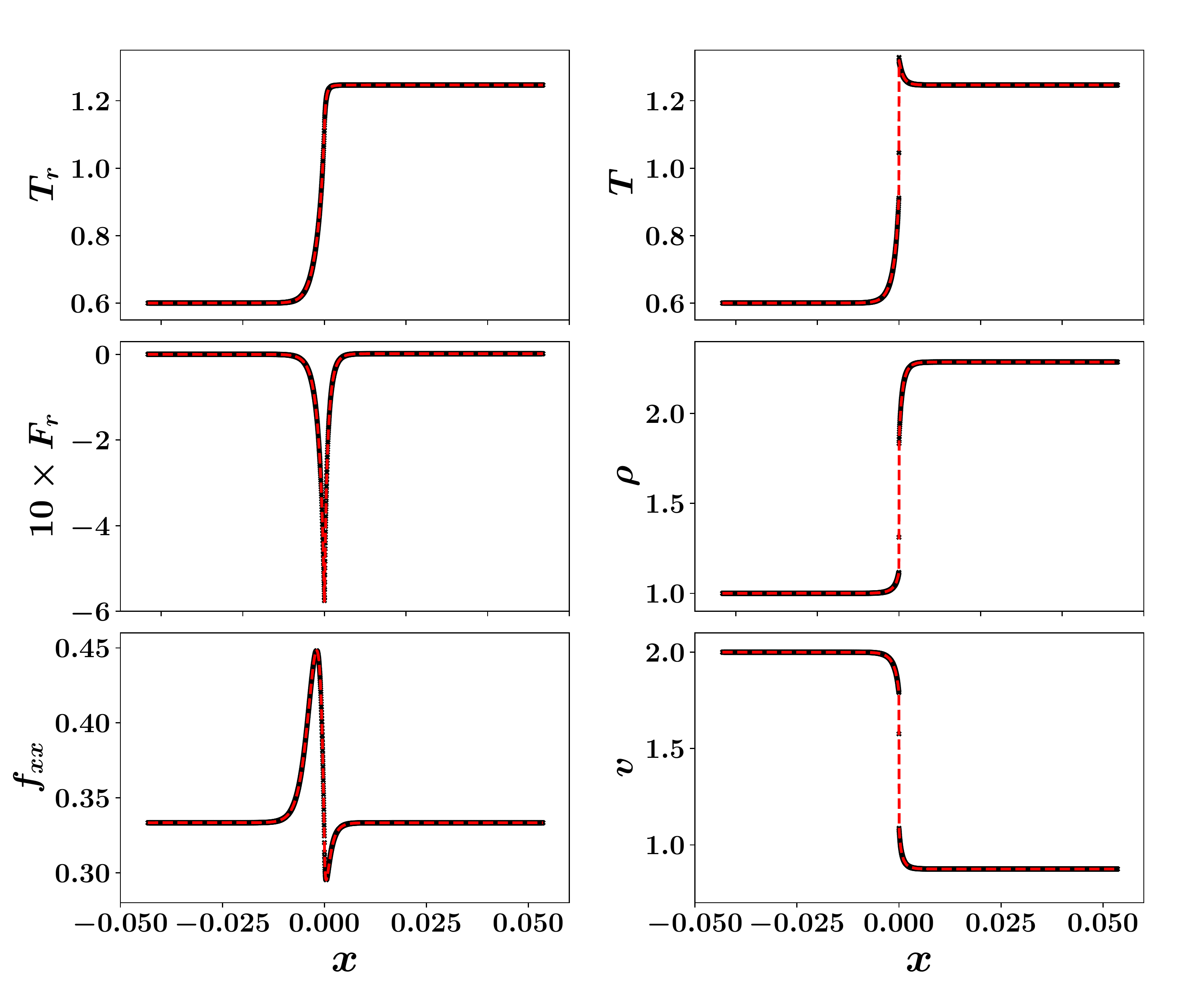}
\caption{Structures of the sub-critical radiation modified shock with upstream Mach number $\mathcal{M}=2$. The panels show steady state profiles of 
radiation temperature $T_r$, gas temperature $T$, radiation flux $F_r$, density $\rho$, horizontal component of Eddington tensor $f_{xx}$ and flow velocity $v$. 
The black lines are from our numerical solution in steady state while the dashed red lines are semi-analytical solutions as described in section \ref{sec:rad_shock}. }
\label{radshock_mp2}
\end{figure}

\begin{figure}[htp]
\centering
\includegraphics[width=1.0\hsize]{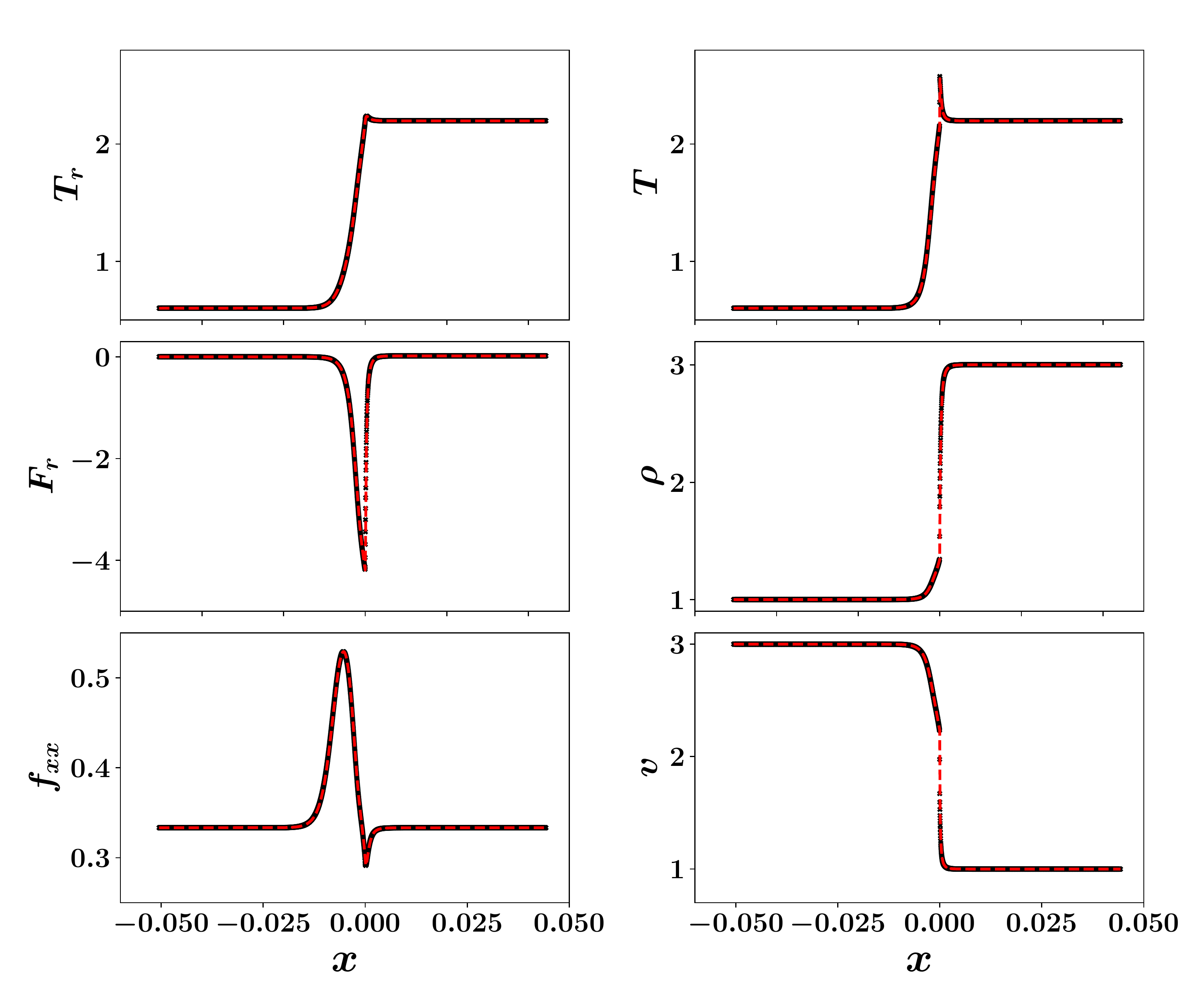}
\caption{The same as Figure \ref{radshock_mp2} but for the case of super-critical shock with Mach number $\mathcal{M}=3$. }
\label{radshock_mp3}
\end{figure}

We choose a unit system such that the upstream adiabatic sound speed is $1$ and the dimensionless speed of light $\Crat=1.73\times 10^3$. The upstream gas has density and temperature $\rho_0=1$ and $T_0=0.6$ with adiabatic index $\gamma=5/3$. We change upstream flow velocity for cases with different Mach numbers. The dimensionless parameter $\Prat$ is $7.72\times 10^{-4}$, which is roughly the ratio between radiation pressure and gas pressure in the upstream. We choose a constant absorption opacity $\rho\kappa_a=577.4$ for the whole simulation domain. We initialize the 1D simulation domain with the semi-analytical solutions given by \cite{Fergusonetal2017} and see how good the code can keep the shock structures. 
We use $4096$ grid points for all the tests. For the left boundary, all the gas quantities are fixed to be the upstream values while the specific intensities are isotropic in the co-moving frame with mean energy density fixed to be $a_rT_0^4$. For the right boundary, we simply copy the gas quantities from the last active zone to the ghost zones but we find it is necessary to keep the gradients of specific intensities to be continuous across the boundary. 

Profiles of various quantities after the numerical solution reaches steady state are shown in Figure \ref{radshock_mp2} for the case with upstream flow velocity $v=2$. This is the case with a sub-critical shock \citep{MihalasMihalas1984,Sincelletal1999,LowrieEdwards2008}, where the gas temperature has a Zel'dovich  spike \citep{ZeldovichRaizer1967} while the radiation temperature is continuous. For the sub-critical case, the gas temperature before the spike is smaller than the downstream temperature. Particularly, the Eddington tensor is smaller than $1/3$ at the right hand side of the spike and larger than $1/3$ at the left hand side of the spike. Our numerical solution in steady state is identical to the semi-analytical solution as shown as the dashed red lines. Notice that our time step is set by the flow speed and we only need roughly $20$ iterations per time step. The numerical cost is significantly smaller than the explicit scheme in \cite{Jiangetal2014} for the same test and it is also more efficient than the VET scheme for a similar test done by \cite{Jiangetal2012}.
The steady state numerical solution with upstream flow velocity increased to $v=3$ is shown in Figure \ref{radshock_mp3}. This is the case with a super-critical shock, which means the gas temperature at the upstream side of the spike is the same as the downstream temperature.  The downstream gas is hotter with an increased ratio between radiation pressure and gas pressure. Our algorithm is still able to keep the shock structures very well as demonstrated in Figure \ref{radshock_mp3} by comparing our numerical solution with the semi-analytical solution.

\subsection{Radiation Driven Wind}
\label{sec:dust_wind}
The simulation of radiation driven dusty wind mentioned in the introduction is a useful test to demonstrate the difference between different approaches to solve the RT equation. This is a test that requires solving the full radiation hydrodynamic equations and the simulation results can be different significantly if the RT equation is not solved accurately. In order to compare with the results based on VET method directly, we use the same setup as in the run $\text{T}3\_\text{F}0.5$ done by \cite{Davisetal2014}. The simulation box is a 2D cartesian domain covering the region $[-256,256]L_0\times[0,1024]L_0$, where the length scale $L_0=2.01\times 10^{15}\text{cm}$ is the fiducial scale height. The spatial resolution is $512\times 1024$ and we use 40 angles per cell.  The fiducial temperature is chosen to be $T_0=82\text{K}$ with the corresponding fiducial velocity 
$v_0=5.4\times 10^4~\text{cm}/\text{s}$ and the fiducial time scale $t_0=3.72\times 10^{10}~\text{s}$. We adopt the same Rosseland and Planck mean dust opacities as in \cite{Davisetal2014}:
\begin{eqnarray}
\kappa_a&=&0.0316\left(\frac{T}{10\text{K}}\right)^2~\text{cm}^2~\text{g}^{-1},\nonumber \\
\kappa_{aP}&=&0.1\left(\frac{T}{10\text{K}}\right)^2~\text{cm}^2~\text{g}^{-1}.
\end{eqnarray}
Notice that $\kappa_{\delta P}=\kappa_{aP}-\kappa_a$ is what we need in equation \ref{eq:cm_source}.
We initialize the gas with a constant density $\rho_0=7.02\times 10^{-16}~\text{g}~\text{cm}^{-3}$ for $9<y/L_0<10$ and density floor $10^{-8}\rho_0$ for all the other regions. The gas has a uniform temperature $T=T_0$ and a constant gravitational acceleration $g=1.45\times 10^{-6}~\text{cm}~\text{s}^{-2}$ along the $-y$ direction. The radiation field is initialized to give $E_r=a_rT_0^4$ and $F_r=0.5c^2g/\kappa_a(T_0)$ through the whole simulation box. We assume upward specific intensities have the same value at each cell and the same assumption is made for downward specific intensities. The same approach is used to set the bottom boundary condition with the same $F_r$ but $E_r=9.26a_rT_0^4$. This is designed to maintain a constant vertical radiation flux $F_r$ with the initial total optical depth $\tau_0=3$. At the top boundary, we set incoming specific intensities to be $0$ and just copy the outgoing specific intensities from the last activate zones to the ghost zones. For gas quantities, we use reflecting boundary condition at the bottom and outflow boundary condition at the top. We use random perturbation on density with amplitude $12.5\%$ to seed the instability. The ratio between speed of light and the fiducial sound speed is $\Crat=5.54\times 10^5$ for this test. 

\begin{figure}[htp]
\centering
\includegraphics[width=1\hsize]{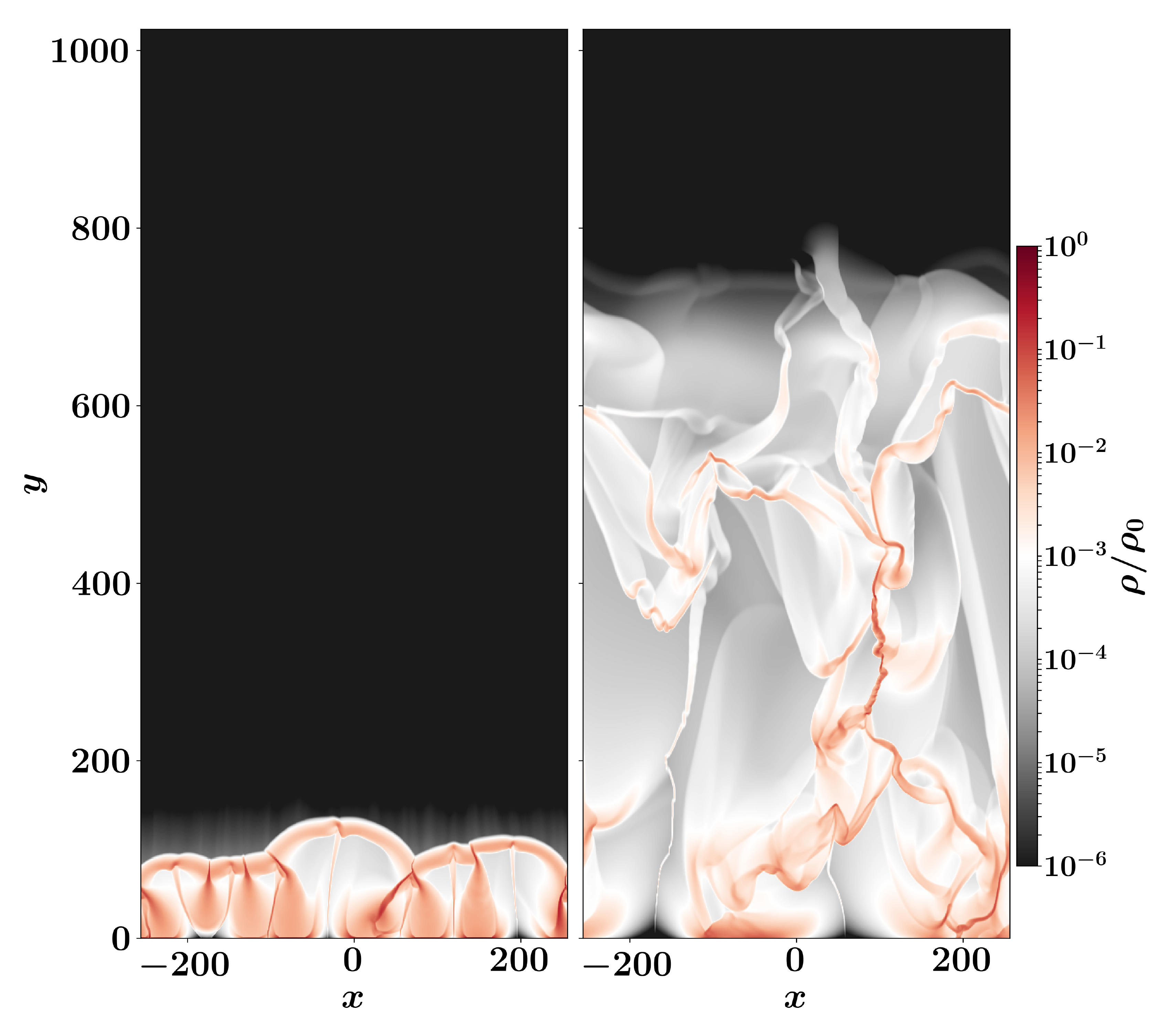}
\caption{Density distribution for two snapshots from the test of radiation driven dusty wind as described in section \ref{sec:dust_wind}. 
The left panel shows the development of radiation driven Rayleigh-Taylor instability at time $t=35t_0$, while the right panel shows the 
fully non-linear stage at time $t=80t_0$. Here $t_0$ is the typical sound cross time across one scale height. }
\label{fig:rad_wind}
\end{figure}

\begin{figure}[htp]
\centering
\includegraphics[width=1\hsize]{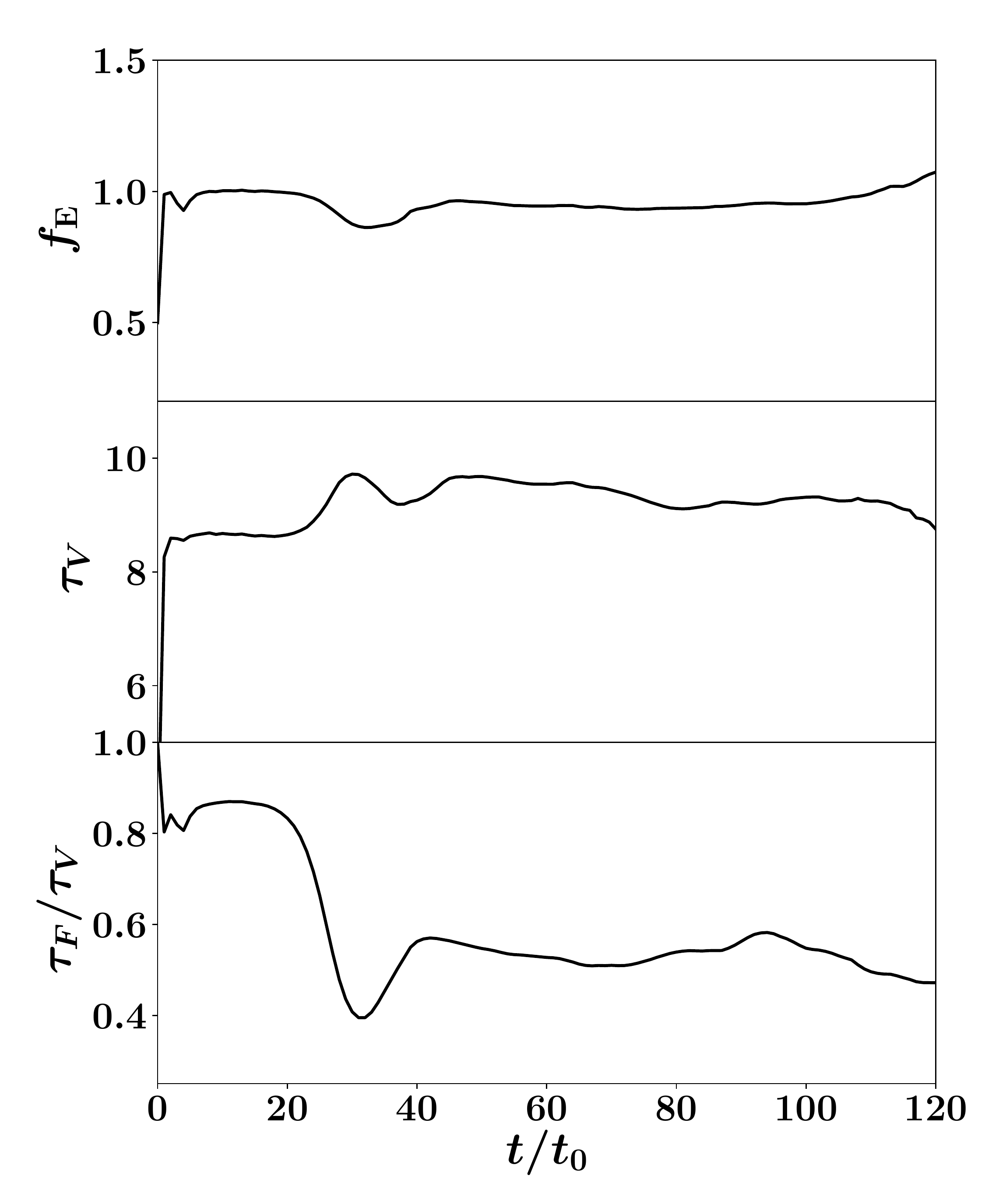}
\caption{Histories of various volume averaged quantities from the test of radiation driven dusty wind (section \ref{sec:dust_wind}). The top panel is 
 the volume averaged Eddington ratio $f_{E}$ while the Rosseland mean optical depth across the horizontal direction of the simulation box is shown in the middle panel. 
 The bottom panel shows the ratio between the flux weighted $(\tau_F)$ and volume averaged $(\tau_V)$ optical depth.}
\label{fig:rad_wind2}
\end{figure}

Density distribution at two snapshots $t=35t_0$ and $80t_0$ are shown in Figure \ref{fig:rad_wind}. The gas shell is pushed away by the radiation force initially and Rayleigh-Taylor instability has developed at $t=35t_0$. The high density filaments fall back temporarily but they get accelerated again later on. At time $t=80t_0$, the majority of the gas is still flowing outward. At each snapshot, we calculate the box averaged Eddington ratio $f_E$, volume $(\tau_V)$ and flux weighted $(\tau_F)$ optical depths as
\begin{eqnarray}
f_E &=& \frac{\langle \rho \kappa_a F_r\rangle}{cg\langle\rho\rangle},\nonumber\\
\tau_V&=&1000L_0\langle\kappa_a\rho\rangle,\nonumber\\
\tau_F&=&1000L_0\frac{\langle \kappa_a\rho F_{r,y}\rangle}{\langle F_{r,y}\rangle},
\end{eqnarray}
where $\langle\cdot\rangle$ represents volume average over the whole simulation box.
Histories of $f_E$, $\tau_V$ and $\tau_F/\tau_V$ are shown in Figure \ref{fig:rad_wind2}. The Eddington ratio dips below 1 around $30t_0$ due to the Rayleigh-Taylor instability but returns to $1$ quickly. The timescale when the dip shows up and the evolution history are basically the same as Figure 6 in \cite{Davisetal2014} despite some small difference at the early phase, which is likely due to slightly different initial condition. This demonstrates that the two completely different algorithms capture the same growth rate and nonlinear structure of the radiation driven Rayleigh-Taylor instability. The Eddington ratio in our simulation eventually becomes larger than 1 and the gas is blown away from the top boundary. The 2D density structures shown in Figure \ref{fig:rad_wind} can also be directly compared with Figure 4 in \cite{Davisetal2014} and they share very similar structures.  

It is also interesting to compare the performance of our algorithm for this test with the performance of VET scheme. The time step for our algorithm is $6\times 10^{-3}t_0$ and we use $50$ iterations per time step to achieve a relative accuracy of $10^{-6}$ except the first few time steps, where $200$ iterations are used per time step to pass the initial transient. With 256 MPI ranks using skylake nodes, we are able to run roughly $2\times 10^4$ steps to reach $120t_0$ within about four hours of wall clock time. The cost per time step is comparable to the cost of the VET method used by \cite{Davisetal2014}. However, \cite{Davisetal2014} used a time step that was a factor of 10 smaller to enable convergence, which caused the overall cost significantly larger than the cost of our new scheme developed here. Our implicit scheme also reduces the cost by a factor of $\approx 10^4$ if we would solve the problem using explicit scheme by limiting the time step with the speed of light.

\section{Discussions and Conclusions}
\label{sec:discuss}
We have developed a novel finite volume algorithm that solves the full time dependent radiation magneto-hydrodynamic equations based on the discrete ordinates approach. 
The hydrodynamic equations are solved using the explicit MHD integrator in {\sf Athena++} while the RT equation is solved fully implicitly. The gas and radiation are coupled via both energy and momentum exchange. The velocity dependent terms in the RT equation are handled via Lorentz transformation so that no expansion in terms of $v/c$ is needed. The time step of the whole algorithm is only determined by the normal CFL condition for the MHD module. We have carried out an extensive set of tests to demonstrate the accuracy and efficiency of the algorithm for both gas pressure and radiation pressure dominated flows with a wide range of optical depth. The finite volume approach also allows the algorithm to be easily used in both Cartesian and Spherical polar coordinate systems as demonstrated in our tests. 

The difference between the algorithm developed here and radiation hydrodynamic schemes that adopt simplified assumption of closure relations for the radiation moment equations (such as FLD and M1) are clear. We do not make any assumptions on the closure relation as we do not need it. Therefore, we do not subject to the well known artifacts of FLD and M1, although our algorithm is generally more expensive. Our algorithm also subjects to the well known ray effects as many other methods based on discrete ordinates \citep{Castor2004}. 
The algorithm is also different from the classical short characteristic method as well as the VET approach. We are solving the full time dependent transport equation for specific intensities directly, instead of the time independent transport equation, which is normally adopted for short characteristic method. It is often argued that  for non-relativistic systems, the light crossing time is much shorter than the local dynamical time scale and therefore it is reasonable to drop the time dependent term for the radiation field. This is widely used to estimate the heating and cooling rate of the gas for the modeling of stellar atmosphere. This is true for the low optical depth regime with negligible radiation pressure. However, when the thermalization time scale is reduced with increasing optical depth, this approach can be numerically unstable even for gas pressure dominated regime \citep{Davisetal2012}. There are also systems where the typical sound speed is much smaller than the speed of light while radiation pressure is still larger than the gas pressure, such as the envelope of massive stars. Our algorithm can overcome the limitations of the short characteristic method and it can be suitable for all these regimes. The VET method can also be used in principle, where a short characteristic module is only used to provide a closure relation. As discussed in Section \ref{sec:shadow}, there is still a subtle difference compared with the algorithm developed here in that the VET approach calculates the Eddington tensor at the lab frame but neglecting the flow velocity. The Eddingon tensor is also calculated at the beginning of each time step  assuming that the radiation field has reached steady state with the gas. Future studies will be needed to investigate the consequences of these differences for applications. Another advantage of our algorithm compared with VET is that the radiation field is entirely described one set of variables (the specific intensities), while VET and Monte Carlo based closure schemes typically need to evolve two completely different sets of radiation variables to get the Eddington tensor and determine the time 
evolution of the radiation moments. It can be an issue to keep the two sets of variables to be consistent with each other \citep{Parketal2012}, which we do not need to worry about.  
The algorithm we have developed here shares many common features with the neutrino transport scheme developed by \cite{SumiyoshiYamada2012}  and \cite{Nagakuraetal2014}. The main differences are the modified HLLE solver we have developed here and the approach to solve the radiation source terms, where we do not linearize the equations. 

The computational cost of this algorithm can be very problem dependent, because the convergent properties can change dramatically depending on the opacity and how quickly the radiation field is changing within one time step. The cost to perform one iteration is smaller than the cost of one step for the explicit scheme. The overall cost is roughly proportional to the number of iterations we need for convergence. The algorithm also shows very good parallel scaling with domain decomposition because unlike the short characteristic module, there is no requirement to perform the iteration with any particular order. This means that each domain can perform the iterations independently and they only need to exchange boundary condition after each iteration. However, MPI communications can take a larger fraction of the total computational cost compared with the explicit scheme.
Because the equations for iteration in our scheme are nonlinear, many pre-conditioners and more efficient matrix inversion algorithms that are designed for linear systems cannot be used here. However, multigrid can be easily compatible with our algorithm and has been shown to be able to speed up the convergence of iterations even for nonlinear equations significantly \citep{Fabianietal1997,Bjorgenetal2017}. Incorporating the multigrid solver in {\sf Athena++} with our RT module is our immediate next step. 

The current algorithm solves the frequency integrated transport equation as the first step. It can be easily extended to the frequency dependent case using the multi-group approach \citep{Vaytetetal2011}, which has been widely used to study neutrino transport \citep{Justetal2015,Skinneretal2019}. The key for this extension is to choose appropriate frequency bins, which can be in the lab frame or the fluid rest frame, or both \citep{Nagakuraetal2014}. All these options will be explored with future development.

\section*{Acknowledgements}
The author thanks James Stone, Shane Davis, Zhaohuan Zhu and the anonymous referee for valuable comments 
that have improved the draft. 
The Center for Computational Astrophysics at the Flatiron Institute 
is supported by the Simons Foundation. 

\bibliographystyle{aasjournal}
\bibliography{citations}

\begin{appendix}
\section{Coefficients of the Discretized RT Equation}
\label{appendix:coef}
In our algorithm, the RT equation is solved fully implicitly as (equation \ref{eqn:In} and section \ref{sec:transport})
\begin{eqnarray}
I_n^{m+1}&+&\Delta t\bfnabla\cdot\left[
\left(c\bn-f\bv^m\right)I_n^{m+1}\right]=
I_n^m-\Delta t\bfnabla\cdot\left(f\bv^mI_n^m\right)\nonumber\\
&+&\Delta tc\Gamma_n^{-3}\left[
\rho^m\kappa_s\left(J_0^{m+1}-I_{0,n}^{m+1}\right)
+\rho^m\kappa_a\left(\frac{a_r\left(T^{m+1}\right)^4}{4\pi}-I_{0,n}^{m+1}\right)
+\rho^m\kappa_{\delta P}\left(
\frac{a_r\left(T^{m+1}\right)^4}{4\pi}-J_0^{m+1}\right)\right].
\label{eq:original}
\end{eqnarray}
At each cell $(i,j,k)$, the left hand side is calculated as
\begin{eqnarray}
I_n^{m+1}&+&\frac{\Delta t}{V_{i,j,k}}\left[
A_x(i+1/2)F_n(i+1/2)-A_x(i-1/2)F_n(i-1/2) \right.\nonumber\\
&+& A_y(j+1/2)F_n(j+1/2)-A_y(j-1/2)F_n(j-1/2) \nonumber\\
&+& \left. A_z(k+1/2)F_n(k+1/2)-A_z(k-1/2)F_n(k-1/2)
\right],
\end{eqnarray}
where $A_x,A_y,A_z$ are the cell areas at the corresponding surfaces while $V_{i,j,k}$ is the cell volume. 
The surface flux is calculated via the modified HLLE solver (taking $F_n(i-1/2)$ as an example)
\begin{eqnarray}
F_n(i-1/2)&=&\frac{S_{i-1/2}^+}{S_{i-1/2}^+-S_{i-1/2}^-}\left[c\mu_x - f v_x^m(i-1/2)\right]I_n^{m+1}(i-1)-
\frac{S_{i-1/2}^-}{S_{i-1/2}^+-S_{i-1/2}^-}\left[c\mu_x - f v_x^m(i-1/2)\right]I_n^{m+1}(i)\nonumber\\
&+&\frac{S_{i-1/2}^+S_{i-1/2}^-}{S_{i-1/2}^+-S_{i-1/2}^-}\left[I_n^{m+1}(i)-I_n^{m+1}(i-1)\right].
\end{eqnarray}
Here the maximum ($S_{i-1/2}^+$) and minimum ($S_{i-1/2}^-$) signal speeds are defined as
\begin{eqnarray}
S_{i-1/2}^+=
\begin{cases}
      c\mu_x\sqrt{\left[1-\exp\left(-\tau_c^2\right)\right]/\tau_c^2}  &  \text{if $\mu_x > 0$}  \\
     -c\mu_x\sqrt{\left[1-\exp\left(-\tau_c^4\right)\right]/\tau_c^2} & \text{if $\mu_x < 0$ }, 
\end{cases}
\end{eqnarray}

\begin{eqnarray}
S_{i-1/2}^-=
\begin{cases}      
     -c\mu_x\sqrt{\left[1-\exp\left(-\tau_c^4\right)\right]/\tau_c^2} & \text{if $\mu_x > 0$ } \\
       c\mu_x\sqrt{\left[1-\exp\left(-\tau_c^2\right)\right]/\tau_c^2}  &  \text{if $\mu_x < 0$}.
\end{cases}
\end{eqnarray}
The optical depth per cell is defined as $\tau_c\equiv \alpha \left[\rho^m(i-1)+\rho^m(i)\right]\left[\kappa_a(i-1)+\kappa_a(i)+\kappa_s(i-1)+\kappa_s(i)\right] \Delta x$, 
where $\Delta x$ is the local cell size and $\alpha$ is a free parameter with the default value chosen to be $5$. The value of $\alpha$ is chosen to minimize numerical diffusion while still keep the numerical scheme stable. For all the test problems we have shown, the solution is independent of $\alpha$ as long as it is large enough. The signal speeds as well as the 
HLLE flux are chosen to preserve the asymptotic limit such that when $\tau_c\rightarrow \infty$, $S_{i-1/2}^+\rightarrow c|\mu_x|/\tau_c$, $S_{i-1/2}^-\rightarrow -c|\mu_x|/\tau_c$ 
and 
\begin{eqnarray}
F_n(i-1/2)\rightarrow \frac{1}{2}\left[c\mu_x-f v^m_x(i-1/2)\right]\left[I_n^{m+1}(i-1)+I_n^{m+1}(i)\right]
-\frac{c|\mu_x|}{2\tau_c}\left[I_n^{m+1}(i)-I^{m+1}_n(i-1)\right].
\end{eqnarray}
In the optically thin limit such that $\tau_c\rightarrow 0$, $F_n(i-1/2)$ is reduced to the simple upwind flux. 
Similarly, surface fluxes from other directions as well the signal speeds $S^+_{j-1/2},S^-_{j-1/2},S^+_{j+1/2},S^-_{j+1/2},S^+_{k-1/2},S^-_{k-1/2},S^+_{k+1/2},S^-_{k+1/2}$ can be calculated.

Replacing the expressions for the fluxes at cell faces, we get the discretized equation
\begin{eqnarray}
& ~&g_1 I_n^{m+1}+g_2I_n^{m+1}(i-1)+g_3I_n^{m+1}(i+1)
+g_4I_n^{m+1}(j-1)\nonumber\\
&+&g_5I_n^{m+1}(j+1)
+g_6I_n^{m+1}(k-1)+g_7I_n^{m+1}(k+1)\nonumber\\
&=&I_n^m-\Delta t\bfnabla\cdot\left(f\bv^mI_n^m\right) \nonumber\\
&+&\Delta tc\Gamma_n^{-3}\left[
\rho^m\kappa_s\left(J_0^{m+1}-I_{0,n}^{m+1}\right)
+\rho^m\kappa_a\left(\frac{a_r\left(T^{m+1}\right)^4}{4\pi}-I_{0,n}^{m+1}\right)
+\rho^m\kappa_{\delta P}\left(
\frac{a_r\left(T^{m+1}\right)^4}{4\pi}-J_0^{m+1}\right)\right],
\end{eqnarray}
where the coefficients are given by
\begin{eqnarray}
g_1&=&1-\frac{A_x(i-1/2) \Delta t}{V_{i,j,k}}\frac{S_{i-1/2}^-}{S_{i-1/2}^+-S_{i-1/2}^-}\left[-c\mu_x + fv_x^m(i-1/2)+S_{i-1/2}^+\right]\nonumber\\
&+&\frac{A_x(i+1/2) \Delta t}{V_{i,j,k}}\frac{S_{i+1/2}^+}{S_{i+1/2}^+-S_{i+1/2}^-}\left[c\mu_x - fv_x^m(i+1/2)- S_{i+1/2}^-\right]\nonumber\\
&-&\frac{A_y(j-1/2) \Delta t}{V_{i,j,k}}\frac{S_{j-1/2}^-}{S_{j-1/2}^+-S_{j-1/2}^-}\left[-c\mu_y + fv_y^m(j-1/2)+S_{j-1/2}^+\right]\nonumber\\
&+&\frac{A_y(j+1/2) \Delta t}{V_{i,j,k}}\frac{S_{j+1/2}^+}{S_{j+1/2}^+-S_{j+1/2}^-}\left[c\mu_y - fv_y^m(j+1/2)- S_{j+1/2}^-\right]\nonumber\\
&-&\frac{A_z(k-1/2) \Delta t}{V_{i,j,k}}\frac{S_{k-1/2}^-}{S_{k-1/2}^+-S_{k-1/2}^-}\left[-c\mu_z + fv_z^m(k-1/2)+S_{k-1/2}^+\right]\nonumber\\
&+&\frac{A_z(k+1/2) \Delta t}{V_{i,j,k}}\frac{S_{k+1/2}^+}{S_{k+1/2}^+-S_{k+1/2}^-}\left[c\mu_z - fv_z^m(k+1/2)- S_{k+1/2}^-\right],\nonumber\\
g_2&=&-\frac{A_x(i-1/2) \Delta t}{V_{i,j,k}}\frac{S_{i-1/2}^+}{S_{i-1/2}^+-S_{i-1/2}^-}\left[c\mu_x - fv_x^m(i-1/2)-S_{i-1/2}^-\right],\nonumber\\
g_3&=&\frac{A_x(i+1/2) \Delta t}{V_{i,j,k}}\frac{S_{i+1/2}^-}{S_{i+1/2}^+-S_{i+1/2}^-}\left[-c\mu_x + fv_x^m(i+1/2)+S_{i+1/2}^+\right],\nonumber\\
g_4&=&-\frac{A_y(j-1/2) \Delta t}{V_{i,j,k}}\frac{S_{j-1/2}^+}{S_{j-1/2}^+-S_{j-1/2}^-}\left[c\mu_y - fv_y^m(j-1/2)-S_{j-1/2}^-\right],\nonumber\\
g_5&=&\frac{A_y(j+1/2) \Delta t}{V_{i,j,k}}\frac{S_{j+1/2}^-}{S_{j+1/2}^+-S_{j+1/2}^-}\left[-c\mu_y + fv_y^m(j+1/2)+S_{j+1/2}^+\right],\nonumber\\
g_6&=&-\frac{A_z(k-1/2) \Delta t}{V_{i,j,k}}\frac{S_{k-1/2}^+}{S_{k-1/2}^+-S_{k-1/2}^-}\left[c\mu_z - fv_z^m(k-1/2)-S_{k-1/2}^-\right],\nonumber\\
g_7&=&\frac{A_z(k+1/2) \Delta t}{V_{i,j,k}}\frac{S_{k+1/2}^-}{S_{k+1/2}^+-S_{k+1/2}^-}\left[-c\mu_z + fv_z^m(k+1/2)+S_{k+1/2}^+\right].
\end{eqnarray}

The convergent properties for our iterative scheme depends strongly on the values of $g_1$. If $g_1$ is large and positive for all $I_n^{m+1}$, the diagonal term will dominate over 
the off-diagonal terms and the iteration can converge quickly. Otherwise, the convergence may be slow and it can even diverge sometimes. To improve the robustness of our iterative scheme, we have an alternative procedure to perform the iteration compared with equation \ref{eqn:iteration}. The expression for $g_1$ includes terms that can be positive or negative depending on the sign of $\mu_x,\mu_y,\mu_z$. For each angle, we separate $g_1$ into two parts $g_1=g_1^{\prime}+g_1^{\prime\prime}$, where $g_1^{\prime}$ includes all the terms that are positive while all the negative terms are included in $g_1^{\prime\prime}$. At each iterative step $l$, we replace $g_1I_{n,l}^{m+1}$ in equation \ref{eqn:iteration} with $g_1^{\prime}I_{n,l}^{m+1}+g_1^{\prime\prime}I_{n,l-1}^{m+1}$. All the other terms are unchanged. This iterative scheme is much more robust over a wide range of parameters we have explored. However, when both schemes can converge, the convergent rate for this procedure is typically slower compared with iteration using equation \ref{eqn:iteration}.

\end{appendix}

\end{CJK*}

\end{document}